%% file: main.tex
\documentclass[final,finalrunningheads,a4paper]{llncs}
\input{styles.tex}
\input{macro.tex}

\newtoggle{anonymous}
\makeatletter
\renewcommand\paragraph{\@startsection{paragraph}{4}{\z@}%
  {2.25ex \@plus 1ex \@minus .2ex}%
  {-0.75em}%
  {\normalfont\normalsize\bfseries}}
\makeatother

\begin{document}

\title{SoK: Lending Pools in Decentralized Finance}

\iftoggle{anonymous}{}{
\author{Massimo Bartoletti\inst{1}, James Hsin-yu Chiang\inst{2}, Alberto Lluch-Lafuente\inst{2}}
% \authorrunning{Bartoletti et al.}

\institute{
Universit\`a degli Studi di Cagliari, Cagliari, Italy
\and 
Technical University of Denmark, DTU Compute, Copenhagen, Denmark}
}

\maketitle

\input{abstract.tex}

\input{introduction.tex}
\input{defi.tex}

\input{lending-pools.tex}
\input{lp-properties.tex}
\input{lp-attacks.tex}

% \input{basic_amm.tex}
% \input{basic_margin.tex}
% \input{ext_model.tex}

\input{related.tex}
\input{conclusion.tex}

\bibliographystyle{splncs04}
\bibliography{main}

\newpage
\appendix
\input{proofs.tex}

\end{document}

%% file: styles.tex
\usepackage[utf8]{inputenc}
\usepackage{color}
\usepackage[usenames,dvipsnames]{xcolor}

\usepackage{centernot}
\usepackage{hhline}
\usepackage{mdframed}

\usepackage{bm} % for bold symbols

\usepackage{mathbbol} % for bb-style greek letters

\usepackage{siunitx}
\usepackage{textcomp}
\usepackage{caption}
\usepackage{pst-func}
\usepackage{amssymb}
\usepackage{graphicx}
\usepackage{paralist}
\usepackage{environ}
\usepackage{pgfplots}
\usepackage{tikz}
\usetikzlibrary{shadows}
\usepackage{mathtools}
\usepackage{array,multirow}
\usepackage{algorithm}
\usepackage[hyphens]{url}
\usepackage{cite}
\usepackage{pgfplotstable}
\usepackage{threeparttable}

\usepackage{etoolbox}
\usepackage{arydshln}

\usepackage{xspace} % Smart spaces after commands
 
\usepackage{dirtytalk} % for \say command
\usepackage{stmaryrd} % for llbracket and rrbracket\textbf{\textbf{}}

\usepackage[inline,shortlabels]{enumitem} 
\newlist{inlinelist}{enumerate*}{1}
\setlist*[inlinelist,1]{%
  label=(\roman*),
}

\usepackage{xifthen}  % Extended conditional commands
\usepackage{ifthen}

\usepackage{nicefrac}

\usepackage[hidelinks]{hyperref}
\usepackage{breakurl} % for arXiv
\usepackage{cleveref}
\usepackage{bigints}

\usetikzlibrary{pgfplots.dateplot}
\usetikzlibrary{matrix,chains,positioning,decorations.pathreplacing}
\usetikzlibrary{patterns,calc,fit,arrows}
\usetikzlibrary{shapes.geometric}

\usepackage[final,nomargin,inline,index]{fixme} % Simplified management of FIXME's
\fxusetheme{color}

\FXRegisterAuthor{bart}{anbart}{\color{magenta} {\underline{bart}}}
\FXRegisterAuthor{james}{anjames}{\color{blue} {\underline{james}}}
\FXRegisterAuthor{alberto}{analberto}{\color{ForestGreen} {\underline{alberto}}}

\hypersetup{
  breaklinks   = true,
  colorlinks   = true, %Colours links instead of ugly boxes
  urlcolor     = blue, %Colour for external hyperlinks
  linkcolor    = blue, %Colour of internal links
  citecolor    = red   %Colour of citations
}
\hypersetup{final} % Needed to generate hyperlinks even in draft mode (ERROR IN MAC)

\usepackage{listings}
\definecolor{listingBG}{HTML}{FFFFCB}%
\definecolor{listingFrame}{HTML}{BBBB98}%
\definecolor{listingLineno}{rgb}{0.5,0.5,1.0}%
\definecolor{LightGrey}{rgb}{0.975,0.975,0.975}
\lstdefinelanguage{balzac}{
	commentstyle=\color{Gray},
	morecomment=[l]{//},
	morecomment=[s]{/*}{*/},
	classoffset=0,
        escapechar=\$,
	morekeywords={const,transaction,input,output,absLock,relLock,key,network,package},
	keywordstyle=\color{TealBlue}\bfseries,
	classoffset=1,
	morekeywords={sig,versig,ctxo,rtxo,ptxo,stxo,verscr,verctx,txid,fun,unit,int,string,bool,address,uint,date,checkDate,sha256},
	keywordstyle=\color{Blue}\bfseries,
	classoffset=2,
	morekeywords={BTC,true,false,and,or,if,then,else},
	keywordstyle=\color{Plum}\bfseries,
	classoffset=3,
	morekeywords={arg,val,wit,out,in,tkid,op,owner,tkval},
	keywordstyle=\color{MidnightBlue}\bfseries,
}

%% file: macro.tex
\newcommand{\ifempty}[3]{%
  \ifthenelse{\isempty{#1}}{#2}{#3}%
}

\newcommand{\ifdots}[3]{%
  \ifthenelse{\equal{#1}{...}}{#2}{#3}%
}

\newcommand{\hidden}[1]{}

\newcommand*\circled[1]{\tikz[baseline=(char.base)]{
            \node[shape=circle,draw,inner sep=0.9pt] (char) {\scriptsize{#1}};}}

\newcommand{\mypar}[1]{\vspace{-0.2cm}\paragraph*{#1}}

\newcommand{\Real}[1]{\mathrm{Real}}

\newcommand{\eg}{e.g.\@\xspace}
\newcommand{\ie}{i.e.\@\xspace}
\newcommand{\wrt}{w.r.t.\@\xspace}

\newcommand{\fst}{\mathit{fst}}
\newcommand{\snd}{\mathit{snd}}

\renewcommand{\epsilon}{\varepsilon}

\newenvironment{proofof}[2][]{%
  \ifempty{#1}
  {\subsection*{Proof of~\Cref{#2}}}
  {\subsection*{Proof of~\Cref{#2} ({#1})}}
  \label{#2-proof}
  }%
  {}

\newcommand{\BTC}{\textup{%
  \leavevmode
  \vtop{\offinterlineskip %\bfseries
    \setbox0=\hbox{B}%
    \setbox2=\hbox to\wd0{\hfil\hskip-.03em
    \vrule height .3ex width .15ex\hskip .08em
    \vrule height .3ex width .15ex\hfil}
    \vbox{\copy2\box0}\box2}}\xspace}

\def\pmvColor{\color{MidnightBlue}}

\newcommand{\pmvFmt}[1]{{\pmvColor{\sf{#1}}}} % active account formatting
\newcommand{\PmvU}{\pmvFmt{\mathbb{A}}\xspace} % universe of ALL user accounts
\newcommand{\pmv}[2][]{\pmvFmt{#2}_{\pmvColor{#1}}\xspace}

\newcommand{\pmvA}[1][]{\pmv[{#1}]{A}}
\newcommand{\pmvAi}[1][]{\pmv[{#1}]{A'}}
\newcommand{\pmvB}[1][]{\pmv[{#1}]{B}}
\newcommand{\pmvC}[1][]{\pmv[{#1}]{C}}
\newcommand{\nrLoanVal}[1]{V^{\textit{nrl}}_{#1}} 
\newcommand{\cMin}{\mathit{C}_{\mathit{min}}}
\newcommand{\rLiq}{\mathit{r}_{liq}}

\newcommand{\supply}[1]{\mathit{sply}_{#1}}

\DeclareMathAlphabet{\mathbfsf}{\encodingdefault}{\sfdefault}{bx}{n}

\newcommand{\irule}[2]{\dfrac{#1}{#2}}

\newcommand{\nrule}[1]{{\scriptsize \textsc{#1}}}

\newcommand{\dom}[1]{\operatorname{dom} {#1}}

\newcommand{\QNN}{\mathbb{Q}^{+}}

\newcommand{\bind}[2]{\nicefrac{#2}{#1}}
\newcommand{\setenum}[1]{\{#1\}}
\newcommand{\setcomp}[2]{\left\{{#1} \,\middle|\, {#2}\right\}}
\crefname{appendix}{appendix}{appendices}
\Crefname{appendix}{Appendix}{Appendices}
\crefname{notation}{notation}{notations}
\Crefname{notation}{Notation}{Notations}

\definecolor{LightGrey}{rgb}{0.95,0.95,0.95}
\definecolor{keyword}{HTML}{7F0055}

\def\tokColor{\color{magenta}}
\newcommand{\tokFmt}[1]{{\tokColor{#1}}}

\newcommand{\tokT}[1][]{\tokFmt{\tau_{#1}}}
\newcommand{\tokTi}[1][]{\tokFmt{\tau'_{#1}}}

\newcommand{\TokU}[1][]{\tokFmt{\mathbb{T}_{#1}}} % universe of all tokens
\newcommand{\TokUF}{\tokFmt{\TokU[\sf f]}} % subset of free tokens
\newcommand{\accrueIntOp}{{\sf Int}} % s-step repay of s-borrowed tokens
\newcommand{\exchUpdateOp}{{\sf Px}}

\newcommand{\ltsLabel}[1][]{{\ell_{#1}}}

\newcommand{\depositOp}{{\sf Dep}}

\newcommand{\redeemOp}{{\sf Rdm}}
\newcommand{\transferOp}{{\sf Trf}} % token transfer
\newcommand{\transferMOp}{{\sf Mtrf}} % co-token transfer
\newcommand{\borrowOp}{{\sf Bor}} % loan borrow
\newcommand{\repayOp}{{\sf Rep}} % loan repay
\newcommand{\liquidateOp}{{\sf Liq}} % (partial) account liquidation
\newcommand{\actDeposit}[2]{\depositOp_{#1}({#2})}

\newcommand{\actRedeem}[2]{\redeemOp_{#1}({#2})}
\newcommand{\actTransfer}[3]{\transferOp_{#1}({#2},{#3})} % token transfer
\newcommand{\actTransferM}[3]{\transferMOp_{#1}({#2},{#3})} % co-token transfer
\newcommand{\actBorrow}[2]{\borrowOp_{#1}({#2})}
\newcommand{\actRepay}[2]{\repayOp_{#1}({#2})}

\newcommand{\actLiquidate}[4]{\liquidateOp_{#1}({#2},{#3},{#4})}

\newcommand{\tokBal}[1][]{\sigma_{#1}} % balance
\newcommand{\tokBali}[1][]{\sigma'_{#1}} % balance
\newlength\replength
\newcommand\repfrac{.1}

\newcommand\rulewidth{.6pt}
\newcommand\tdashfill[1][\repfrac]{\cleaders\hbox to \replength{%
  \smash{\rule[\arraystretch\ht\strutbox]{\repfrac\replength}{\rulewidth}}}\hfill}

\newcommand\tdotfill[1][\repfrac]{\cleaders\hbox to \replength{%
  \smash{\raisebox{\arraystretch\dimexpr\ht\strutbox-.1ex\relax}{.}}}\hfill}

\newcommand{\val}[2][]{#2_{#1}} % values
\newcommand{\valV}[1][]{\val[#1]{v}}
\newcommand{\valVi}[1][]{\val[#1]{v'}}
\newcommand{\valVii}[1][]{\val[#1]{v''}}

\newcommand{\ER}[2][]{\mathit{ER}_{#1}\ifempty{#2}{}{({#2})}} % Exchange Rate
\newcommand{\Coll}[1]{\mathit{C}_{#1}} % Account Collateralization
\newcommand{\Util}[1]{\mathit{U}_{#1}} % Exchange Rate
\newcommand{\Intr}[1]{\mathit{I}_{#1}} % Exchange Rate

\newcommand{\uTok}[1]{\mathit{u}_{#1}} 

\newcommand{\minted}[1][]{\TokU[{#1}]}

\newcommand{\confG}[1][]{\Gamma_{#1}}

\newcommand{\lpFfun}{\pi_{f}}
\newcommand{\lpFfuni}{\pi'_{f}}

\newcommand{\lpMfun}{\pi_{m}}
\newcommand{\lpMfuni}{\pi'_{m}}

\newcommand{\lpBfun}{\pi_{l}}
\newcommand{\lpBfuni}{\pi'_{l}}

\newcommand{\LpS}[1][]{\pi_{#1}} % overall lending pool state
\newcommand{\LpSi}[1][]{\pi'_{#1}}

\newcommand{\exchO}{{\it p}}
\newcommand{\exchOi}{{\it p}'}

\newcommand{\collVal}[1][]{V^m_{#1}}
\newcommand{\loanVal}[1][]{V^l_{#1}}

%% file: abstract.tex
\begin{abstract}
  Lending pools are decentralized applications which allow mutually untrusted 
  users to lend and borrow crypto-assets. 
  These applications feature complex, highly parametric incentive mechanisms 
  to equilibrate the loan market.
  This complexity makes the behaviour of lending pools difficult to understand and to predict:
  indeed, ineffective incentives and attacks could potentially lead to emergent unwanted behaviours.
  Reasoning about lending pools is made even harder
  by the lack of executable models of their behaviour:
  to precisely understand how users interact with lending pools, 
  eventually one has to inspect their implementations, 
  where the incentive mechanisms are intertwined with low-level implementation details. 
  Further, the variety of existing implementations makes it difficult 
  to distill the common aspects of lending pools.
  We systematize the existing knowledge about lending pools, leveraging
  a new formal model of interactions with users, which reflects the archetypal features of mainstream implementations.
  This enables us to prove some general properties of lending pools, 
  such as the correct handling of funds, 
  and to precisely describe vulnerabilities and attacks.
  We also discuss the role of lending pools in the broader context 
  of decentralized finance.
\end{abstract}

%% file: introduction.tex
% \albertonote{We need a better, more informative title. Some options that come to my mind, but we shall discuss (it depends also on our main message and the final content):}
% \begin{itemize}
% \item SoK: Models, properties and vulnerabilities of DeFi smart contracts
% \item SoK: Models, properties and vulnerabilities of smart contracts for decentralized finance
% \item SoK: DeFi smart contracts, formally
% \item SoK: A formalisation of DeFi smart contracts
% \item SoK: Formalisation and vulnerabilities of DeFi smart contracts
% \item SoK: (variants of the above with ``Decentralized Finance'' expanded)
% \item SoK: (variants of the above ``DeFi smart contracts archetypes'' or some other word)
% \end{itemize}

\section{Introduction} \label{section:introduction}

The emergence of permissionless, public blockchains has given birth to an 
entire ecosystem of \emph{crypto-tokens} representing digital assets.
Facilitated and accelerated by smart contracts and standardized token 
interfaces \cite{ERC20}, these so-called \emph{decentralized finance} (DeFi) applications 
promise an open alternative to the traditional financial system.
One of the main DeFi applications are \emph{lending pools}, 
which incentivize users to lend some of their crypto-assets to borrowers.
Unlike in traditional finance, all the parameters of a loan, like its interests, maturity periods or token prices, are determined by a smart contract, which also includes mechanisms to incentivize honest behaviour (\eg, loans are eventually repaid), economic growth and stability. 
Existing lending pool platforms are already handling large volumes of crypto-assets: 
as of writing, the two main platforms % , Compound and Aave, 
currently hold \$1.7B \cite{compstats} and \$1.4B \cite{aavestats} 
worth of tokens in their smart contracts.

%Leading automatic market makers Uniswap and Curve Finance 
%hold \$1.6B \cite{uniswapstats} and \$1.5B \cite{curvestats} worth of tokens and feature %\$320M \cite{uniswapstats} and \$36M \cite{curvestats} 
%worth of swap transactions every day. 
%\albertonote{Again... should we avoid mentioning AMMs. Otherwise we could create an expectation that does not become true :)}

Lending pools are inherently hard to design.
Besides the typical difficulty of implementing secure smart contracts
\cite{ABC17post,DAO,parity17jul,parity17nov},
lending pools feature complex economic incentive mechanisms, 
which make it difficult to understand when 
a lending pool actually achieves the economic goals it was designed for.
As a matter of fact, a recent failure of the oracle price feed used by 
the Compound lending pool platform led to \$100M of collateral being 
(incorrectly) liquidated \cite{compoundoracleattack}.
Indeed, most current literature in DeFi is devoted to study 
the economic impact of these incentive mechanisms~\cite{gudgeon2020decentralized,gudgeon2020plf,perez2020liquidations, kaoanalysis,chitra2019competitive,chitra2020stake}.
% As a matter of fact, 
% subtle vulnerabilities in the economic incentives of lending pools
% and in their actual implementations 
% were recently discovered~\cite{gudgeon2020decentralized,kaoanalysis}. 
% As a matter of fact, 
% recent analyses of lending pool 
% safety~\cite{gudgeon2020decentralized,kaoanalysis} 
% must rely on sophisticated assumptions about agent behaviour 
% and interactions with external markets.

The problem is made even more complex by the absence of 
abstract operational descriptions of the behaviour of lending pools.
Current descriptions are either high-level economic models~\cite{gudgeon2020plf, gudgeon2020decentralized, perez2020liquidations}, 
or the actual implementations.
While, on the one hand, 
economic models are useful to understand the macroscopic financial aspects of lending pools,
on the other hand they do not precisely describe the interactions between a lending pool and its users.
Still, understanding these interactions is crucial to determine if a lending pool is vulnerable to attacks where some users deviate from the expected behaviour. 
Implementations, instead, reflect the exact actual behaviour, but at a level of detail that makes high-level understanding and reasoning unfeasible.

\paragraph{Contributions} 

This paper presents a systematic analysis of the behaviour of lending pools, 
of their properties, vulnerabilities, and of the related literature.
Based on a throughout inspection of the implementations of the two main 
lending pool platforms, 
Compound \cite{compimpl} and Aave \cite{aaveimpl},
we synthesise a formal, operational model of the interactions between users and lending pools,
encompassing their incentive mechanisms.
More specifically, our contributions are: 
\begin{enumerate}

\item a formal model of lending pools, 
  which precisely describes their interactions 
  as transitions of a state machine. 
  Our model captures all the typical transactions of lending pools, 
  and all the main economic features, like collateralization, exchange rates, 
  token price, and interest accrual (\Cref{sec:lending-pools});
  
\item the formalization and proof of fundamental behavioural properties 
  of lending pools,
  which were informally stated in literature,
  and are expected to be satisfied by any implementation
  (\Cref{sec:lp-properties});
  % We prove that these properties hold in our formal model;
  % We define and prove properties which guarantee the safety of lent 
  % and borrowed funds under specific conditions;
  
\item the formalization of relevant properties of the incentive mechanisms
  of lending pools, and a discussion of their vulnerabilities and attacks
  (\Cref{sec:lp-attacks});
  % The safety of lending pools rely on the assumptions that price volatility 
  % is bounded and that the incentivisation for users to repay loans 
  % which have not been sufficiently maintained is effective (liquidation). 
  % Here, we define safety properties which indicate whether aforementioned 
  % assumptions hold in a given LP state.

\item a thorough discussion on the interplay between lending pools and other
  DeFi archetypes, like stable coins and automatic market makers 
  (\Cref{sec:related}).

\end{enumerate}

Overall, our contributions help address the aforementioned challenges 
in the design of lending pools.
Firstly, our formal model provides a precise understanding of the behaviour
of lending pools, abstracting from low-level implementation details.
Our model is faithful to mainstream  lending pool implementations
like Compound \cite{compimpl} and Aave \cite{aaveimpl};
still, for the sake of clarity, we have introduced high-level abstractions 
over low-level details:
we discuss the differences between our model and the actual lending pool 
platforms in~\Cref{sec:conclusions}.
Secondly, our formalisation of the properties of the 
incentive mechanisms of lending pools  
makes it easier to understand and analyse their vulnerabilities and attacks. 
In this regard, our model is directly amenable 
for its interpretation as an \emph{executable specification}, 
thus paving the way for automated analysis techniques, 
which may include mechanised proofs of contract properties 
and agent-based simulations of lending pools and other DeFi contracts.

% \paragraph{Structure of the paper}

% Section~\ref{section:Background} provides a gentle introduction to DeFi.
% Section~\ref{sec:basic-blockchain} introduces the basic model
% Sections~\ref{sec:lending-pools} introduces the model of LPs. 
% We exploit the model to formalize properties of LPs 
% and illustrate possible attacks and vulnerabilities 
% in~\Cref{sec:lp-properties,sec:lp-attacks}.
% Section~\ref{sec:related} provides an outlook on DeFi smart contracts, including a discussion of the role of Lending Pools. 
% Section~\ref{sec:conclusion} concludes the paper with a discussion of the main differences of our model with respect to LP implementations and other models, and outlines future research avenues.

%% file: defi.tex
\section{Background} \label{section:Background}

% We gently introduce the behaviour of lending pools,
% financial concepts relevant to lending pools, 
% but which are also featured in other DeFi archetypes, 
% such as algorithmic stable coins and automatic market makers. 

Lending pools (in short, LPs) are financial applications which 
% interconnect lenders and borrowers, 
create a market of loans of crypto-assets, 
providing incentive mechanisms to equilibrate the market.
We now overview the main features of LPs;
a glossary of LP terms is in~\Cref{table: Financial terms}. 

Users can lend assets to a LP
by transferring \emph{tokens} from their accounts to the LP. 
In return, they receive a \emph{claim}, 
represented as tokens \emph{minted} by the LP,
which can later be redeemed for an equal or increased amount 
of tokens, of the same \emph{token type} of the original deposit. 
Lending is incentivized by interest or fees:
the depositor speculates that the claim will be redeemable 
for a value greater than that of the original deposit.
% The depositor receives minted tokens in exchange 
% for the deposit of tokens. 
% promises a future exchange for the previously deposited token, 
Users can redeem claims by transferring minted tokens to the LP,
which pays back the original tokens (with accrued interest) to the redeemer, 
simultaneously burning the minted tokens.
However, redeeming claims is not always possible, as the LP could not have
a sufficient balance of the original tokens, 
as these may have been lent to other users. 
% In order for a minted token to be fungible, 
% the exchange rate between minted and deposited tokens 
% must be consistent for deposits and redeems.

User initiate a \emph{loan} by borrowing tokens deposited to a LP.
To incentivise users to eventually repay the loan,
borrowing requires to provide a \emph{collateral}. 
Collaterals can be either tokens deposited to the LP
when the loan is initiated, and locked for the whole loan duration,
or they can be tokens held by the borrower but \emph{seizable} by the LP
when a user fails to repay a loan. 
% Since LPs cannot spontaneously trigger actions, 
% LPs incentivize users to seize collaterals: namely, 
An unpaid loan of $\pmvA$ can be \emph{liquidated} by $\pmvB$,
who pays (part of) $\pmvA$'s loan in return for a discounted amount 
of $\pmvA$'s collateral. 
For this to be possible, the value of the collateral must be greater  
than that of the loan. 
% (\ie, the loan is \emph{over-collateralized}). 
To incentivize deposits, loans \emph{accrue} interest, 
which increase a user's loan amount by the \emph{interest rate}. 

\begin{table}[h!]
  \centering
  \scriptsize
  \begingroup
  \renewcommand{\arraystretch}{1.3} % Default value: 1
  \begin{tabular}{l  p{9cm}}\hline
    \textbf{Token} & A digital representation of some asset, transferable between users. 
    \\
    \textbf{Token type} & A set of tokens.  
    Tokens of a given type are interchangeable (or \emph{fungible}), 
    whereas tokens of different token types are not. 
    \\
    \textbf{Native token} & The default token type of a blockchain (\eg, ETH for Ethereum). 
    \\
    \textbf{Token price} & The price of a token type $\tokT$ 
    is the amount of units of a given native crypto-currency (or fiat currency) needed to buy one unit of $\tokT$.
    % exchange rate $\tokT / \tokT[n]$ between a token type $\tokT \in \TokU$ and the native token type $\tokT[n]$. The price of $\tokT$ is defined by an external market.
    \\
    \textbf{Exchange rate} & Given two token types $\tokT$ and $\tokTi$, the ratio $\tokT / \tokTi$ at which a user can exchange units of token type $\tokTi$ for units of $\tokT$ in a blockchain interaction.
    \\
%    \textbf{Arbitrage} & A user-blockchain interaction, where a user can make a risk-free profit. 
%    \\
    \textbf{Lender} & A user who transfers units of a token type in return for a \emph{claim} on a full repayment in the future, which may include additional fees or interest. 
    \\
    \textbf{Claim} & A right to token units in the future. Claims are represented as tokens, which are \emph{minted} and destroyed as claims are created and redeemed. 
    \\
    \textbf{Minting} & Creation of tokens performed by the LP upon deposits.
    \\
    \textbf{Borrower} & A user who wishes to obtain a \emph{loan} of token type $\tokT$. The borrower is required to hold \emph{collateral} of another token $\tokTi$ to secure the loan.
    \\
    \textbf{Collateral} & A user balance of tokens which can be seized if the user does not adequately repay a loan.
    \\
    \textbf{Collateralization} & The ratio of deposited \emph{collateral} value over the borrower's total loan value. 
    \\
    \textbf{Liquidation} & When the \emph{collateralization} of user $\pmvA$ falls below a minimum threshold it is \emph{undercollateralized}: here, a user $\pmvB$ can repay a fraction of $\pmvA$'s loan, in return for a discounted amount of $\pmvA$'s collateral \emph{seized} by $\pmvB$. 
    \\
    \textbf{Interest rate} & The rate of loan growth when accruing interest.
    \\ 
    \hline
  \end{tabular}
  \endgroup
  \caption{Glossary of financial terms used in Lending Pools.}
  \label{table: Financial terms}
\end{table}

%% file: lending-pools.tex
\section{Lending pools}
\label{sec:lending-pools}

In this~\namecref{sec:lending-pools}
we introduce a formal, operational model of lending pools.
We do this incrementally, starting from a basic model of blockchains, 
on top of which we will specify the behaviour lending pools.

\subsection{A basic model of blockchains} 
\label{sec:basic-blockchain}

% \begin{table}[h]
%   \centering
%   \begin{tabular}{ll}
%     \hline
%     \begin{tabular}{ll}
%       $\pmvA,\pmvAi,\ldots \in \PmvU$ 
%       & users 
%       \\
%       $\tokT,\tokTi,\ldots \in \TokU$ 
%       & token types 
%       \\
%       $\tokBal \in \PmvU \rightharpoonup (\TokU \rightharpoonup \Nat)$
%       & user token balance \\
%       $\tokBal[\pmvA]$ 
%       & abbreviation of $\tokBal \pmvA$ 
%       \\
%     \end{tabular}
%       & \begin{tabular}{ll}
%           $\exchO \in \TokU[f] \rightarrow \Nat$ 
%           & price of free tokens
%           \\
%           $\confG, \confGi$ 
%           & blockchain state
%           \\
%           $\ell$ 
%           & action 
%           \\[-3pt]
%           $\confG\xrightarrow{\ell} \confGi$ 
%           & state transition
%         \end{tabular}
%     \\
%     \hline
%   \end{tabular}
%   \caption{Summary of notation.}
%   \label{table: notations - blockchain model}
% \end{table}

We assume a set of \emph{users} $\PmvU$, ranged over by $\pmvA, \pmvAi, \ldots$,
and a set of \emph{token types} $\TokU$, ranged over by $\tokT,\tokTi,\ldots$.
We denote with $\TokU[f] \subseteq \TokU$ the subset of 
tokens types that can be freely transferred between users,
only assuming a sufficient balance of the sender
($\TokU[f]$ includes \eg the native blockchain tokens).
% , like bitcoins and ether).

We render blockchain states as partial maps
$\tokBal \in \PmvU \rightharpoonup (\TokU \rightharpoonup \QNN)$,
where $\tokBal \pmvA$ represents $\pmvA$'s token balance
(a partial map from token types to nonnegative rational numbers).
% and $\br \in \Nat^{+}$ is a blockchain round,
% representing the passing of time 
% (\eg, the blockheight in the Ethereum blockchain). 
% updated periodically, such as the blockheight in the Ethereum blockchain: It's update can be interpreted as the passing of a fixed time period.
Hereafter, we abbreviate $\tokBal \pmvA$ as $\tokBal[\pmvA]$. 
We use the standard notation $f \setenum{\bind{x}{v}}$ 
to update a partial map $f$ at point $x$:
namely, $f \setenum{\bind{x}{v}}(x) = v$, while
$f \setenum{\bind{x}{v}}(y) = f(y)$ for $y \neq x$.

Given a partial map $f \in \TokU \rightharpoonup \QNN$, 
a token type $\tokT \in \TokU$ and a partial binary operation  
$\circ \in \QNN \times \QNN \rightharpoonup \QNN$, 
we define the partial map $f \circ \valV:\tokT$ as follows:
\begin{equation} \label{eq: balance update}
  f \circ \valV:\tokT = \begin{cases}
    f\setenum{\bind{\tokT}{f(\tokT) \;\circ\; \valV}}
    & \text{if $\tokT \in \dom{f}$ and $f(\tokT) \circ \valV$ is defined} \\
    %f\setenum{\bind{\tokT}{0}} & \text{if $\tokT \in \dom{f}$ and $\valVi \circ \valV$ is undefined} 
    %\\
    f\setenum{\bind{\tokT}{\valV}}
    & \text{if $\tokT \not\in \dom{f}$}
  \end{cases}
\end{equation}

\noindent
% For instance, 
We adopt the notation $\valV:\tokT$ to denote $\valV$ units of token $\tokT$ throughout the paper.

We model the interaction between users and the blockchain 
as a state transition system,
with labels $\ltsLabel$ which represent transactions.
Our basic model has only one kind of transaction,
$\actTransfer{\pmvA}{\pmvB}{\valV:\tokT}$,
which represents the transfer of $\valV:\tokT$ from $\pmvA$ to $\pmvB$. 
Its effect on the state is specified by the following rule:
\begin{equation*} \label{eq:Trf}
  \small
  \irule
  {\begin{array}{cccc}
     \circled{1}\; \tokBal[\pmvA](\tokT) \geq \valV \quad\;
     & \circled{2}\; \tokT \in \TokU[f] \quad\;
     & \circled{3}\; \tokBali[\pmvA] = \tokBal[\pmvA] - \valV:\tokT \quad\;
     & \circled{4}\; \tokBali[\pmvB] = \tokBal[\pmvB] + \valV:\tokT
   \end{array}}
 {
   \tokBal
   \xrightarrow{\actTransfer{\pmvA}{\pmvB}{\valV:\tokT}}
   \tokBal \setenum{\bind{\pmvA}{\tokBali[\pmvA]}} \setenum{\bind{\pmvB}{\tokBali[\pmvB]}} 
 }
 \;\nrule{[Trf]}
\end{equation*}
We decorate rule preconditions with circled numbers, \eg \circled{1}, 
to simplify their reference in the text. 
Rule \nrule{[Trf]} states that the transfer is permitted whenever
the sender has a sufficient balance \circled{1}, 
and the transferred token type is free \circled{2}.

\subsection{Lending pool states} \label{section:LPstate}

We now extend our basic blockchain model with lending pools,
focussing on the common features implemented by the main platforms.
We make our model parametric \wrt platform-specific features, 
like \eg interest rate models, and we abstract from some advanced features,
like \eg governance (see \Cref{sec:conclusions} for
a discussion on the differences between our model and the existing platforms).

% Hereafter we denote with $\TokUM = \TokU \setminus \TokUF$ the 
% set of tokens \emph{minted} by the LP.

We model states $\confG$ as terms of the form
$\tokBal \mid \LpS \mid \exchO$, where
$\tokBal$ is the token balance of users,
$\LpS$ is the lending pool state,
and  $\exchO \in \TokU[f] \rightarrow \QNN$
models an oracle who prices the free tokens.
Lending pool states $\LpS$ are triples $(\lpFfun,\lpBfun,\lpMfun)$, where:
\begin{itemize}
\item $\lpFfun \in \TokUF \rightharpoonup \QNN$
  records the balance of free token types deposited in the LP;
\item $\lpBfun \in \PmvU \rightharpoonup (\TokUF \rightharpoonup \QNN)$ 
  records the amount and type of tokens lent to users; 
\item $\lpMfun \in \TokUF \rightharpoonup ((\TokU \setminus \TokUF) \times \QNN)$ 
  records the amount of tokens minted by the LP upon deposits. 
  Namely, $\lpMfun(\tokT) = (\tokTi,n)$ means that 
  the LP owns $n$ units of a token type $\tokTi$ 
  minted to represent claims of deposited tokens of type $\tokT$. 
  We require that different free tokens are associated to different 
  minted tokens: 
  \begin{equation}
    \label{eq:lpm:inj}
    \lpMfun(\tokT[1]) = (\tokTi,n_1)
    \;\land\;
    \lpMfun(\tokT[2]) = (\tokTi,n_2)
    \;\implies\; 
    \tokT[1] = \tokT[2]
  \end{equation}
\end{itemize}

We denote with $\minted[\LpS]$ the set of tokens minted by a LP in state $\LpS$.
For a minted token $\tokTi \in \minted[\LpS]$, 
we denote with $\uTok{\LpS}(\tokTi)$ the \emph{underlying} free token.
Formally:
\begin{equation} \label{eq:uTok}
  \minted[\LpS] = \setcomp{\fst(\lpMfun(\tokT))}{\tokT \in \TokUF}
  \qquad\qquad
  \uTok{\LpS}(\tokTi) = \tokT 
  \quad \text{if $\fst(\lpMfun(\tokT)) = \tokTi$}
\end{equation}

\noindent
Note that~\eqref{eq:lpm:inj} ensures that
$\uTok{\LpS}(\tokTi[1]) \neq \uTok{\LpS}(\tokTi[2])$
when $\tokTi[1] \neq \tokTi[2]$.
We say that a state $\tokBal \mid \LpS \mid \exchO$ 
is \emph{initial} if 
$\lpFfun$, $\lpBfun$, $\lpMfun$ have empty domain,
and $\dom{\tokBal[\pmvA]} \subseteq \TokUF$ for all $\pmvA$.

\begin{table}[t!]
  \centering
  \footnotesize
  \begin{tabular}{ll}
    \hline
    $\actDeposit{\pmvA}{\valV:\tokT}$ 
    & $\pmvA$ deposits $\valV$ units of a free token $\tokT$, 
      receiving minted tokens 
    \\
    $\actBorrow{\pmvA}{\valV:\tokT}$ 
    & $\pmvA$ borrows $\valV$ units of free token $\tokT$ 
    \\
     $\accrueIntOp$ 
    & All loans accrue interest
    \\   
    $\actRepay{\pmvA}{\valV:\tokT}$ 
    &  $\pmvA$ repays $\valV$ units on $\pmvA$'s loan in $\tokT$ 
    \\
    $\actRedeem{\pmvA}{\valV:\tokT}$ 
    & $\pmvA$ redeems $\valV$ units of minted $\tokT$, receives deposited tokens 
    \\
    $\actLiquidate{\pmvA}{\pmvB}{\valV:\tokT}{\valVi:\tokTi}$ 
    & $\pmvA$ repays $\valV$ units of $\pmvB$'s loan in $\tokT$, seizing $\valVi : \tokTi$ from $\pmvB$
    \\
    $\actTransferM{\pmvA}{\pmvB}{\valV:\tokT}$ 
    & $\pmvA$ transfers $\valV$ units of minted $\tokT$ to $\pmvB$
    \\
    $\actTransfer{\pmvA}{\pmvB}{\valV:\tokT}$
    & $\pmvA$ transfers $\valV$ units of free $\tokT$ to $\pmvB$
    \\
    \hline
  \end{tabular}
  \caption{Lending pool actions.}
  \label{table: LP tx types}
  \label{tab:lending-pool:tx}
\end{table}

\begin{table}[t!]
\centering
\scriptsize
\begingroup
\begin{tabular}{|l|c|c|c|c|c|c|c|c|c|c|c|c|c|c|}
\hline
\multirow{2}{*}{Actions} & \multicolumn{4}{c|}{$\tokBal[\pmvA]$} & \multicolumn{3}{c|}{$\tokBal[\pmvB]$} & \multicolumn{2}{c|}{$\lpFfun$} & $\lpBfun \,\pmvB $ & \multicolumn{2}{c|}{$\lpMfun$} & \multicolumn{2}{c|}{$\exchO$} \\ \cline{2-15} 
 & $\tokT[0]$ & $\tokT[1]$ & $\tokTi[0]$ & $\tokTi[1]$ & $\tokT[0]$ & $\tokT[1]$ & $\tokTi[1]$ & $\tokT[0]$ & $\tokT[1]$ & $\tokT[0]$ & $\tokT[0]$ & $\tokT[1]$ & $\tokT[0]$ & $\tokT[1]$ \\ \hline
0. Initial State & 100 & -- & -- & -- & -- & 50 & -- & -- & -- & -- & -- & -- & 1 & 1 \\ \hline
1. $\actDeposit{\pmvA}{50:\tokT[0]}$ & \textbf{50} & -- & \textbf{50} & -- & -- & 50 & -- & 50 & -- & -- & \textbf{$\tokTi[0]$:50} & -- & 1 & 1 \\ \hline
2. $\actDeposit{\pmvB}{50:\tokT[1]}$ & 50 & -- & 50 & -- & -- & 0 & \textbf{50} & 50 & 50 & -- & $\tokTi[0]$:50 & \textbf{$\tokTi[1]$:50} & 1 & 1 \\ \hline
3. $\actBorrow{\pmvB}{30:\tokT[0]}$ & 50 & -- & 50 & -- & \textbf{30} & 0 & 50 & 20 & 50 & \textbf{30} & $\tokTi[0]$:50 & $\tokTi[1]$:50 & 1 & 1 \\ \hline
4. $\accrueIntOp$ & 50 & -- & 50 & -- & 30 & 0 & 50 & 20 & 50 & \textbf{34} & $\tokTi[0]$:50 & $\tokTi[1]$:50 & 1 & 1 \\ \hline
5. $\actRepay{\pmvB}{5:\tokT[0]}$ & 50 & -- & 50 & -- & \textbf{25} & 0 & 50 & \textbf{25} & 50 & \textbf{29} & $\tokTi[0]$:50 & $\tokTi[1]$:50 & 1 & 1 \\ \hline
6. $\exchUpdateOp$ & 50 & -- & 50 & -- & 25 & 0 & 50 & 25 & 50 & 29 & $\tokTi[0]$:50 & $\tokTi[1]$:50 & \textbf{1.3} & 1 \\ \hline
7. $\actLiquidate{\pmvA}{\pmvB}{13:\tokT[0]}{19:\tokTi[1]}$ & \textbf{37} & -- & 50 & \textbf{19} & 25 & 0 & \textbf{31} & \textbf{38} & 50 & \textbf{16} & $\tokTi[0]$:50 & $\tokTi[1]$:50 & 1.3 & 1 \\ \hline
8. $\actRedeem{\pmvA}{10:\tokTi[0]}$ & \textbf{48} & -- & \textbf{40} & 19 & 25 & 0 & 31 & \textbf{27} & 50 & 16 & \textbf{$\tokTi[0]$:40} & $\tokTi[1]$:50 & 1.3 & 1 \\ \hline
\end{tabular}
\endgroup
\caption{Interactions between two users and a lending pool.}
\label{tab:lending-pool:ex1}
\end{table}

\subsection{An overview of lending pools behaviour} 
\label{section:LPappetizer}

Lending pools support several actions, summarized in~\Cref{tab:lending-pool:tx}.
Before formalizing their behaviour, we give some intuition
through an example involving two users $\pmvA$ and $\pmvB$ 
(see~\Cref{tab:lending-pool:ex1}). 
$\pmvA$ and $\pmvB$ start by depositing 50 units of 
free tokens $\tokT[0]$ and $\tokT[1]$,
for which they receive equal amounts of freshly minted tokens
$\tokTi[0]$ and $\tokTi[1]$. 

Next, $\pmvB$ borrows $30:\tokT[0]$. 
%, previously deposited by $\pmvA$. 
Here, the 50 minted tokens of type $\tokTi[1]$ in $\pmvB$'s balance 
serve as \emph{collateral} for the loan. % of 30 units of $\tokT[0]$. 
The \emph{collateralization} of $\pmvB$ is the ratio between 
the \emph{value} of $\pmvB$'s balance of $\tokTi[1]$ 
and the value of $\pmvB$'s loan of $\tokT[0]$ 
(the value of a token balance is the product between the number of units 
of the token and its price).
Assuming a minimum collateralization threshold of $\cMin = 1.5$ 
and equal token prices for $\tokT[0]$ and $\tokT[1]$, 
$\pmvB$ could borrow up to 34 units of $\tokT[1]$, 
given the collateral of $50:\tokTi[1]$. 
Nonetheless, $\pmvB$ decides to leave some margin 
to manage future price volatility and the accrual of interest, 
which can both negatively affect collateralization. 
In action 4, interest accrues on the loan made by $\pmvB$.
Here, the interest 
% accrual 
rate is 12\%, 
so $\pmvB$'s loan amount grows from $30$ to $34$ units of $\tokT[1]$. 
In action 5, $\pmvB$ repays 5 units of $\tokT[0]$ to
reduce the risk of becoming \emph{liquidated}, which can occur when
$\pmvB$'s collateralization falls below the threshold $\cMin = 1.5$.

Despite this effort, the price is updated in action 6, such that $\exchO(\tokT[0])$ 
increases by 30\% relative to $\exchO(\tokT[1])$, thereby decreasing the relative value of 
$\pmvB$'s collateral to $\pmvB$'s loan.
As a result, the collateralization of $\pmvB$ drops
below the threshold $\cMin$. 
In action 7, $\pmvA$ liquidates \mbox{$13:\tokT[0]$} of $\pmvB$'s loan, 
restoring $\pmvB$'s collateralization to $\cMin$, 
and simultaneously seizing \mbox{$19:\tokTi[1]$} from $\pmvB$'s balance.
The exchange of $13:\tokT[0]$ for $19:\tokTi[1]$ implies 
a liquidation discount, which ensures that 
the liquidation is profitable for the user performing it.

In action 8, $\pmvA$ then \emph{redeems} $10: \tokTi[0]$, 
receiving $11:\tokT[0]$ in exchange. 
Here, each unit of $\tokTi[0]$ is now 
exchanged for more than 1 unit of $\tokT[0]$, due to accrued interest.

\subsection{Lending pool transitions} \label{section:LPrules}

We now present the full set of rules which
formalize the behaviour of lending pools.
To illustrate them, we provide an extended running example
(Tables~\ref{table: example LP deposits}--\ref{table: example LP liquidation}).
% with three users $\pmvA$, $\pmvB$, $\pmvC$,
% assuming an initial state $\tokBal \mid \LpS \mid \exchO$,
% where the initial state of the accounts $\tokBal$ is shown 
% in the first row of~\Cref{table: example LP deposits}.  
%
% The price oracle $\exchO$ is left unspecified, 
% unless it plays a role in the rules being illustrated. 

\mypar{Deposit} 

A user $\pmvA$ can deposit $\valV$ units of a token $\tokT$
by performing the transaction $\actDeposit{\pmvA}{\valV:\tokT}$,
provided that the balance is sufficient \circled{1}.
In return, $\pmvA$ receives $\valVi$ units of a token $\tokTi$
minted by the LP. 
Upon the first deposit of $\tokT$, 
the LP creates a fresh (non-free) token type $\tokTi$ \circled{2};
freshness ensures that condition~\eqref{eq:lpm:inj} is
preserved by the new state.
For further deposits of $\tokT$, the LP mints new units of $\tokTi$.
In both cases, the amount of minted units of $\tokTi$ 
are recorded in the $\lpMfun$ component of the state \circled{5}. 
Note that premises \circled{2} and \circled{5}
require that $\tokT$ must be a free token type.
The amount $\valVi$ \circled{3}
is the ratio between the deposited amount $\valV$
and the exchange rate $\ER[\LpS]{\tokT}$ between $\tokT$ and $\tokTi$, 
defined in~\eqref{eq: LP exchange rate}. 
% bart: not needed to anticipate here
% The same exchange rate holds for the execution of \emph{redeem} in 
% \eqref{eq: LP redeem transaction}, where LP-minted tokens are exchanged 
% for units of the corresponding, deposited free token type.
% \jamesnote{Add comment about "fresh"}
\begin{equation*} \label{eq: LP deposit rule}
  \small
  \irule
  {
    \begin{array}{l}
      \circled{1}\; \tokBal[\pmvA](\tokT) \geq \valV 
      % \qquad
      % \circled{2}\; \tokT \in \TokUF 
      \qquad
      \circled{2}\; \tokTi := \begin{cases}
        \text{fresh $\not\in \TokUF$} & \text{if $\tokT \not\in \dom \lpMfun$} \\   
        \lpMfun(\tokT) & \text{otherwise}
      \end{cases}
      \qquad
      \circled{3}\; \valVi := \nicefrac{\valV}{\ER[\LpS]{\tokT}} 
      \\[12pt]
      \circled{4}\; \lpFfuni := \lpFfun + \valV:\tokT 
      \qquad
      \circled{5}\; \lpMfuni := \begin{cases}
        \lpMfun\setenum{\bind{\tokT}{(\tokTi,\valVi)}}
        & \mathit{if} \; \tokT \not\in \dom \lpMfun
        \\
        \lpMfun\setenum{\bind{\tokT}{(\tokTi,\valVii+\valVi)}}
        & \mathit{if} \; \lpMfun(\tokT) = (\tokTi,\valVii)
      \end{cases}
    \end{array}
  }
  {\begin{array}{l}
     \tokBal \mid
     \LpS \mid \exchO
     \xrightarrow{\actDeposit{\pmvA}{\valV:\tokT}}
     \tokBal \setenum{\bind{\pmvA}{\tokBal[\pmvA] \,-\, \valV:\tokT \,+\, \valVi:\tokTi }} \mid
     (\lpFfuni,\lpBfun,\lpMfuni)
     \mid \exchO
   \end{array}
 }
 \;\nrule{[Dep]}
\end{equation*}

% LP-table-deposits
\begin{table}[t!]
\centering
\caption{Running example: deposit actions} 
\scriptsize
\begingroup
\label{table: example LP deposits}
\begin{tabular}{|l|c|c|c|c|c|c|c|c|c|c|c|c|c|c|c|c|}
\hline
\multirow{3}{*}{Actions} & \multicolumn{4}{c|}{\multirow{2}{*}{$\tokBal[\pmvA]$}} & \multicolumn{4}{c|}{\multirow{2}{*}{$\tokBal[\pmvB]$}} & \multicolumn{2}{c|}{\multirow{2}{*}{$\tokBal[\pmvC]$}} & \multicolumn{3}{c|}{\multirow{2}{*}{$\lpFfun$}} & \multicolumn{3}{c|}{\multirow{2}{*}{$\lpMfun$}} \\
 & \multicolumn{4}{c|}{} & \multicolumn{4}{c|}{} & \multicolumn{2}{c|}{} & \multicolumn{3}{c|}{} & \multicolumn{3}{c|}{} \\ \cline{2-17} 
 & $\tokT[0]$ & $\tokT[1]$ & $\tokTi[0]$ & $\tokT[1]$ & $\tokT[0]$ & $\tokT[2]$ & $\tokTi[0]$ & $\tokTi[2]$ & $\tokT[2]$ & $\tokTi[2]$ & $\tokT[0]$ & $\tokT[1]$ & $\tokT[2]$ & $\tokT[0]$ & $\tokT[1]$ & $\tokT[2]$ \\ \hline
0. Initial state & 100 & 300 & - & - & 50 & 50 & - & - & 100 & - & - & - & - & - & - & - \\ \hline
1. $\actDeposit{\pmvA}{100:\tokT[0]}$ & \textbf{0} & 300 & \textbf{100} & - & 50 & 50 & - & - & 100 & - & \textbf{100} & - & - & \textbf{$\tokTi[0]$:100} & - & - \\ \hline
2. $\actDeposit{\pmvA}{150:\tokT[1]}$ & 0 & \textbf{150} & 100 & \textbf{150} & 50 & 50 & - & - & 100 & - & 100 & \textbf{150} & - & $\tokTi[0]$:100 & \textbf{$\tokTi[1]$:150} & - \\ \hline
3. $\actDeposit{\pmvB}{50:\tokT[0]}$ & 0 & 150 & 100 & 150 & \textbf{0} & 50 & \textbf{50} & - & 100 & - & \textbf{150} & 150 & - & \textbf{$\tokTi[0]$:150} & $\tokTi[1]$:150 & - \\ \hline
4. $\actDeposit{\pmvB}{50:\tokT[2]}$ & 0 & 150 & 100 & 150 & 0 & \textbf{0} & 50 & \textbf{50} & \textbf{100} & - & 150 & 150 & \textbf{50} & $\tokTi[0]$:150 & $\tokTi[1]$:150 & \textbf{$\tokTi[2]$:50} \\ \hline
5. $\actDeposit{\pmvC}{100:\tokT[2]}$ & 0 & 150 & 100 & 150 & 0 & 0 & 50 & 50 & \textbf{0} & 100 & 150 & 150 & \textbf{150} & $\tokTi[0]$:150 & $\tokTi[1]$:150 & \textbf{$\tokTi[2]$:50} \\ \hline
\end{tabular}
\endgroup
\end{table}

The main idea of the exchange rate is that, 
while initially there is a 1/1 correspondence between 
minted and deposited tokens, 
when interest is accrued this relation changes to the benefit of lenders. 
For a free token $\tokT$, the exchange rate $\ER[\LpS]{\tokT}$
represents the share of deposited units of $\tokT$ 
over the units of the associated minted tokens. 
If any loans remain pending, not all minted tokens can be redeemed, as only a fraction of the deposited free tokens remain in the LP balance. 
\bartnote{so what?? Lent units of $\tokT$ positively contribute to the exchange rate}
Formally: 
\begin{equation} \label{eq: LP exchange rate}
  \ER[\LpS]{\tokT} = 
  \dfrac
  {
    \lpFfun(\tokT)
    + 
    \sum_{\pmvA} (\lpBfun \, \pmvA) \, \tokT
  }
  {\snd(\lpMfun(\tokT))}
  \; \textit{if $\lpFfun(\tokT) > 0$}
  \qquad
  \ER[\LpS]{\tokT} = 1
  \; \textit{if $\lpFfun(\tokT) = 0$}
\end{equation}
where we assume that the items $\pmvA$ for which 
$\lpBfun \, \pmvA$ or $(\lpBfun \, \pmvA) \tokT$ are undefined
do not contribute to the summation
(we will adopt this convention through the paper).

\Cref{table: example LP deposits} exemplifies users depositing funds to the LP. 
In transaction 1, $\pmvA$ deposits 100 units of $\tokT[0]$.
Since this is the first deposit of $\tokT$, 
the LP mints exactly 100 units of a \emph{fresh} token type, say $\tokTi[0]$, 
and transfers these units to $\pmvA$. 
In transaction 2, $\pmvA$ deposits 150 units of $\tokT[1]$;
similarly to the previous case, $\pmvA$ receives 150 units of 
a fresh token type $\tokTi[1]$.
In transaction 3, $\pmvB$ deposits 50 units of $\tokT[0]$.
Since $\tokT[0]$ was already deposited, 
the LP mints 50 units of the existing token type $\tokTi[0]$, 
and transfers them to $\pmvB$.
%Here, the exchange rate $\tokT[1]/\tokTi[1]$ remains 1.
Finally, in transactions 4 and 5 
$\pmvB$ and $\pmvC$ deposit units of $\tokT[2]$; 
after that, the balances of tokens $\tokT[0]$, $\tokT[1]$, $\tokT[2]$ 
in the LP total 150 units.

\mypar{Borrow} 

Any user can borrow units of a free token type $\tokT$ from the LP,
provided that the LP has a sufficient balance of $\tokT$ \circled{1},
and that the user has enough minted tokens to use as collateral \circled{4}. 
More specifically, we require that the \emph{collateralization} of the user
is above a constant threshold $\cMin > 1$. 
\begin{equation*} \label{eq: LP borrow transaction}
  \small
  \irule
  {
    \begin{array}{ll}
      \circled{1}\; \lpFfun(\tokT) \geq \valV > 0 
      & \circled{2}\; f_{\pmvA} = \begin{cases}
        \lpBfun \pmvA + \valV:\tokT
        & \text{if $\pmvA \in \dom{\lpBfun}$} \\
        \setenum{\bind{\tokT}{\valV}}
        & \text{otherwise}
      \end{cases}
      \\[12pt]
      \circled{3}\; \LpSi := (\lpFfun - \valV:\tokT,\, \lpBfun\setenum{\bind{\pmvA}{f_{\pmvA}}},\, \lpMfun)
      \mbox{\quad}
      & \circled{4}\; \Coll{\tokBal \mid \LpSi \mid \exchO}(\pmvA) \geq \cMin 
    \end{array}
  }
  {
    \begin{array}{l}
      \tokBal \mid \LpS \mid \exchO
      \xrightarrow{\actBorrow{\pmvA}{\valV:\tokT}}
      \tokBal \setenum{\bind{\pmvA}{\tokBal[\pmvA] + \valV:\tokT}} \mid 
      \LpSi \mid 
      \exchO
    \end{array}}
  \;\nrule{[Bor]}
\end{equation*}

\noindent
To define the collateralization of users,
we introduce a few auxiliary notions.
The value $\loanVal(\pmvA)$ of $\pmvA$'s loans is the sum 
(over all free token types $\tokT$) of the \emph{value} of $\tokT$-tokens
lent to $\pmvA$ (the value is the product between token amount and price).
For instance, if $\pmvA$ has borrowed only 10 tokens of type $\tokT$, 
and the price of $1:\tokT$ is $2:\tokT[n]$, then the value of $\pmvA$'s 
loan is $20:\tokT[n]$.
Formally:
\begin{equation} \label{eq:loanVal}
  \loanVal[\confG](\pmvA) 
  \; = \;
  \textstyle
  \sum_{\tokT \in \TokUF} (\lpBfun \, \pmvA) \tokT \cdot \exchO(\tokT)
  \qquad \text{if $\confG = \tokBal \mid \LpS \mid \exchO$}
\end{equation}

The value $\collVal(\pmvA)$ of minted tokens held by $\pmvA$
is the summation (over all minted token types $\tokT$)
of the \emph{value} of $\pmvA$'s balance of minted tokens.
To determine the value of an minted token $\tokTi$, 
its price is equated to that of the 
underlying free token $\tokT$, as minted tokens do not 
exist in the domain of~$\exchO$:
% (\Cref{table: notations - blockchain model}).
\begin{equation} \label{eq:collVal}
  \collVal[\confG](\pmvA)
  \; = \;
  \textstyle \sum_{\tokT \in \TokU \setminus \TokUF}
  \tokBal[\pmvA](\tokT) \cdot \ER[\LpS]{\uTok{\LpS}(\tokT)}
  \cdot \exchO(\uTok{\LpS}(\tokT))
  \qquad \text{if $\confG = \tokBal \mid \LpS \mid \exchO$}
\end{equation}

\noindent
The collateralization of a user 
is the ratio of the value of minted to lent tokens:
\begin{equation} \label{eq: LP account collateralization}
  \Coll{\confG}(\pmvA) 
  \; = \;
  {\collVal[\confG](\pmvA)}
  \; / \;
  {\loanVal[\confG](\pmvA)}
\end{equation}

% \noindent
% In \cref{eq: LP account collateralization}, $\collTok[\LpS](\pmvA)$ and $\borrTok[\LpS](\pmvA)$ denote the set of LP-minted token types in 
% $\pmvA$'s balance and the set of token types
% borrowed by $\pmvA$ respectively.

We exemplify $\borrowOp$ transactions in~\Cref{table: example LP borrowing}. 
Users $\pmvB$ and $\pmvC$ borrow 
amounts of $\tokT[0]$ and $\tokT[1]$ at steps 6--8,
keeping their collateralization above $\cMin$, 
which is assumed to be $1.5$.
% Lent amounts are recorded in $\lpBfun$. 
$\pmvC$'s collateralization decreases from 3.3 to 1.7 upon step 8: 
this is due to the increase in 
$\loanVal(\pmvC)$, whilst $\collVal(\pmvC)$ 
remains constant at $100$.

% LP-table-borrows
\begin{table}[t!]
\centering
\caption{Running example: borrow actions}
\label{table: example LP borrowing}
\scriptsize
\begingroup
\begin{tabular}{|l|c|c|c|c|c|c|c|c|c|c|c|c|c|c|c|c|c|c|c|c|}
\hline
\multirow{3}{*}{Actions} & \multicolumn{5}{c|}{\multirow{2}{*}{$\tokBal[\pmvB]$}} & \multicolumn{4}{c|}{\multirow{2}{*}{$\tokBal[\pmvC]$}} & \multicolumn{3}{c|}{$\lpBfun$} & \multicolumn{3}{c|}{\multirow{2}{*}{$\lpFfun$}} & \multicolumn{3}{c|}{\multirow{2}{*}{$\exchO$}} & \multicolumn{2}{c|}{\multirow{2}{*}{$\Coll{\confG}$}} \\ \cline{11-13}
 & \multicolumn{5}{c|}{} & \multicolumn{4}{c|}{} & $\pmvB$ & \multicolumn{2}{c|}{$\pmvC$} & \multicolumn{3}{c|}{} & \multicolumn{3}{c|}{} & \multicolumn{2}{c|}{} \\ \cline{2-21} 
 & $\tokT[0]$ & $\tokT[1]$ & $\tokT[2]$ & $\tokTi[0]$ & $\tokTi[1]$ & $\tokT[0]$ & $\tokT[1]$ & $\tokT[2]$ & \multicolumn{1}{l|}{$\tokTi[2]$} & $\tokT[1]$ & $\tokT[0]$ & $\tokT[1]$ & $\tokT[0]$ & $\tokT[1]$ & $\tokT[2]$ & $\tokT[0]$ & \multicolumn{1}{l|}{$\tokT[1]$} & \multicolumn{1}{l|}{$\tokT[2]$} & $\pmvB$ & $\pmvC$ \\ \hline
5. $\actDeposit{\pmvC}{100:\tokT[2]}$ & 0 & - & 0 & 50 & 50 & - & - & 0 & 100 & - & - & - & 150 & 150 & \textbf{150} & 1 & 1 & 1 & - & - \\ \hline
6. $\actBorrow{\pmvB}{50:\tokT[1]}$ & 0 & \textbf{50} & 0 & 50 & 50 & - & - & 0 & 100 & \textbf{50} & - & - & 150 & \textbf{100} & 100 & 1 & 1 & 1 & \textbf{2.0} & - \\ \hline
7. $\actBorrow{\pmvC}{30:\tokT[0]}$ & 0 & 50 & 0 & 50 & 50 & \textbf{30} & - & 0 & 100 & 50 & \textbf{30} & - & \textbf{120} & 100 & 100 & 1 & 1 & 1 & 2.0 & \textbf{3.3} \\ \hline
8. $\actBorrow{\pmvC}{30:\tokT[1]}$ & 0 & 50 & 0 & 50 & 50 & 30 & \textbf{30} & 0 & 100 & 50 & 30 & \textbf{30} & 120 & \textbf{70} & 70 & 1 & 1 & 1 & 2.0 & \textbf{1.7} \\ \hline
\end{tabular}
\endgroup
\end{table}

As we have seen, user collateralization depends on the amount of minted tokens he possesses, the amount of tokens, and the price of all tokens involved. Therefore, collateralization is potentially sensitive to all actions 
that can affect those values. 
This includes both interest accrual and 
changes in token prices (which are unpredictable). 
Borrowers must therefore maintain a safety margin 
in order to protect against potential liquidation.
% ~\eqref{eq: LP liquidate transaction}. 

\mypar{Interest Accrual}

Interest accrual models the periodic application of interest to 
loan amounts and can be executed in any state. 
%(unlike other actions, this one is not triggered by users).  
The action applies a token-specific interest $\Intr{\LpS}(\tokT)$
to each loan, updating the $\lpBfun$ mapping for \emph{all} users.
\begin{equation*} \label{eq: LP interest compounding}
  \small
  \irule
  {
    \begin{array}{l}
      \lpBfuni(\pmvA) := f'_{\pmvA}
      \;\text{ if $\pmvA \in \dom \lpBfun$, where}
      \;\;
      f'_{\pmvA}(\tokT) := (\Intr{\LpS}(\tokT) + 1) \cdot (\lpBfun \, \pmvA) \tokT
      \;\text{ if $\tokT \in \dom (\lpBfun \pmvA)$}
    \end{array}
  }
  {\tokBal \mid \LpS \mid \exchO
    \; \xrightarrow{\accrueIntOp} \;
    \tokBal \mid 
    (\lpFfun,\lpBfuni,\lpMfun)
    \mid \exchO
  }
  \;\nrule{[Int]}
\end{equation*}

Existing lending pool platforms deploy different algorithmic 
interest rate models~\cite{gudgeon2020plf}. 
We leave our model parametric \wrt interest rates, 
and only require that the interest rate is positive, 
a property that all models in~\cite{gudgeon2020plf} satisfy:
\begin{equation} \label{eq: interest rate}
\begin{array}{c}
    % \Intr{\LpS}(\tokT) \in \mathbb{R}_{[0,1]} \rightarrow \mathbb{R}_{\geq0}
    % \qquad
    \Intr{\LpS}(\tokT) > 0 \\
\end{array}
\end{equation}

We extend our running example with three interest updates 
in~\Cref{table: example LP interest}, resulting in the increase of all loan
amounts. 
Each subsequent execution of $\accrueIntOp$ \emph{decreases} the  
collateralization of users $\pmvB$ and $\pmvC$, since the $\loanVal$ of
both borrowers \emph{increases} as interest is applied~\eqref{eq: LP account collateralization}.

% LP-table-interest
\begin{table}[t!]
\centering
\caption{Running example: interest accrual}
\label{table: example LP interest}
\scriptsize
\begingroup
\begin{tabular}{|l|c|c|c|c|c|c|c|c|c|c|c|}
\hline
\multirow{3}{*}{Actions} & \multicolumn{3}{c|}{$\lpBfun$} & \multicolumn{3}{c|}{\multirow{2}{*}{$\Intr{\LpS}$}} & \multicolumn{3}{c|}{\multirow{2}{*}{$\exchO$}} & \multicolumn{2}{c|}{\multirow{2}{*}{$\Coll{\confG}$}} \\ \cline{2-4}
 & $\pmvB$ & \multicolumn{2}{c|}{$\pmvC$} & \multicolumn{3}{c|}{} & \multicolumn{3}{c|}{} & \multicolumn{2}{c|}{} \\ \cline{2-12} 
 & $\tokT[1]$ & $\tokT[0]$ & $\tokT[1]$ & $\tokT[0]$ & $\tokT[1]$ & $\tokT[2]$ & $\tokT[0]$ & $\tokT[1]$ & $\tokT[2]$ & $\pmvB$ & $\pmvC$ \\ \hline
8. $\actBorrow{\pmvC}{30:\tokT[1]}$ & 50 & 30 & \textbf{30} & 2.0\% & 5.3\% & 0\% & 1 & 1 & 1 & 2.00 & \textbf{1.67} \\ \hline
9. $\accrueIntOp$ & \textbf{53} & \textbf{31} & \textbf{32} & 2.1\% & 5.5\% & 0\% & 1 & 1 & 1 & \textbf{1.89} & \textbf{1.59} \\ \hline
10. $\accrueIntOp$ & \textbf{56} & \textbf{32} & \textbf{34} & 2.1\% & 5.6\% & 0\% & 1 & 1 & 1 & \textbf{1.79} & \textbf{1.52} \\ \hline
11. $\accrueIntOp$ & \textbf{59} & \textbf{33} & \textbf{36} & 2.2\% & 5.8\% & 0\% & 1 & 1 & 1 & \textbf{1.69} & \textbf{1.45} \\ \hline
\end{tabular}
\endgroup
\end{table}

\mypar{Repay} 

A user with a loan can repay part of it by executing a $\repayOp$ transaction:
% given that he/she has a sufficient 
% balance \circled{1}.
\begin{equation*} \label{eq: LP repay transaction}
  \small
  \irule
  {\begin{array}{l}
     \circled{1}\; \tokBal[\pmvA](\tokT) \geq \valV > 0 
     \qquad
     \circled{2}\; (\lpBfun \, \pmvA) \, \tokT \geq \valV
     \qquad\,
     \circled{3}\; \lpBfuni = \lpBfun \setenum{\bind{\pmvA}{\lpBfun \pmvA - \valV:\tokT}}
   \end{array}}
 {\begin{array}{l}
    \tokBal \mid \LpS \mid \exchO
    \xrightarrow{\actRepay{\pmvA}{\valV:\tokT}}
    \tokBal \setenum{\bind{\pmvA}{\tokBal[\pmvA] - \valV:\tokT}} \mid
    (\lpFfun + \valV:\tokT,\, \lpBfuni,\, \lpMfun) \mid
    \exchO
  \end{array}}
  \;\nrule{[Rep]}
\end{equation*}

\noindent
This increases the collateralization of the repaying user, 
as $\loanVal$ is reduced~\eqref{eq: LP account collateralization}. 
Users must always maintain a sufficient collateralization, 
to cope with adverse effects of interest accruals and price updates.

In \Cref{table: example LP repay}, $\pmvC$ is suffering 
from low collateralization after the last interest accrual in transaction 11. 
Here, $\Coll{\confG}(\pmvC)$ is equal to $\cMin = 1.5$. 
The subsequent repayment of 15 units of $\tokT[0]$
increases $\pmvC$'s collateralization back to 1.9.

% LP-table-repay
\begin{table}[t!]
\centering
\caption{Running example: repay actions}
\label{table: example LP repay}
\scriptsize
\begingroup
\begin{tabular}{|l|c|c|c|c|c|c|c|c|c|c|c|c|c|c|c|c|c|c|c|c|c|}
\hline
\multirow{3}{*}{Actions} & \multicolumn{4}{c|}{\multirow{2}{*}{$\tokBal[\pmvA]$}} & \multicolumn{5}{c|}{\multirow{2}{*}{$\tokBal[\pmvB]$}} & \multicolumn{4}{c|}{\multirow{2}{*}{$\tokBal[\pmvC]$}} & \multicolumn{3}{c|}{\multirow{2}{*}{$\lpFfun$}} & \multicolumn{3}{c|}{$\lpBfun$} & \multicolumn{2}{c|}{\multirow{2}{*}{$\Coll{\confG}$}} \\ \cline{18-20}
 & \multicolumn{4}{c|}{} & \multicolumn{5}{c|}{} & \multicolumn{4}{c|}{} & \multicolumn{3}{c|}{} & $\pmvB$ & \multicolumn{2}{c|}{$\pmvC$} & \multicolumn{2}{c|}{} \\ \cline{2-22} 
 & $\tokT[0]$ & $\tokT[1]$ & $\tokTi[0]$ & $\tokTi[1]$ & $\tokT[0]$ & $\tokT[1]$ & $\tokTi[2]$ & $\tokTi[0]$ & $\tokTi[2]$ & $\tokT[0]$ & $\tokT[1]$ & $\tokT[2]$ & $\tokTi[2]$ & $\tokTi[0]$ & $\tokTi[1]$ & $\tokTi[2]$ & $\tokT[1]$ & $\tokT[0]$ & $\tokT[1]$ & $\pmvB$ & $\pmvC$ \\ \hline
11. $\accrueIntOp$ & 0 & 150 & 100 & 150 & 0 & 50 & 0 & 50 & 50 & 30 & 30 & 0 & 100 & 120 & 70 & 150 & \textbf{59} & \textbf{33} & \textbf{36} & \textbf{1.7} & \textbf{1.5} \\ \hline
12. $\actRepay{\pmvC}{15:\tokT[0]}$ & 0 & 150 & 100 & 150 & 0 & 50 & 0 & 50 & 50 & \textbf{15} & 30 & 0 & 100 & \textbf{135} & 70 & 150 & 59 & \textbf{18} & 36 & 1.7 & \textbf{1.9} \\ \hline
\end{tabular}
\endgroup
\end{table}

\mypar{Redeem} 

A user without any loans can redeem minted tokens $\tokT$ \circled{1} for the 
underlying tokens if enough units of $\uTok{\LpS}(\tokT)$ 
remain in the LP~\circled{2}. 
A user with a \emph{non-zero} loan amount of 
any token can only redeem minted tokens such that the resulting  
collateralization is not below $\cMin$ \circled{3}. 
This constraint does not apply to users without loans, 
as minted tokens are not used as collateral.
\begin{equation*} \label{eq: LP redeem transaction}
  \small
  \irule
  {\begin{array}{l} 
     \circled{1}\; \tokBal[\pmvA](\tokT) \geq \valV > 0 
     \qquad
     \valVi := \valV \cdot \ER[\LpS]{\uTok{\LpS}(\tokT)} 
     \qquad
     \circled{2}\; \lpFfun(\uTok{\LpS}(\tokT)) \geq \valVi 
     \\[4pt]
     \circled{3}\;  (\exists \tokTi. (\lpBfun \pmvA)\tokTi > 0) \Rightarrow \Coll{\tokBali \mid \LpSi \mid \exchO}(\pmvA) \geq \cMin 
     \qquad
     \tokBali[\pmvA] := \tokBal[\pmvA] - \valV:\tokT + \valVi:\uTok{\LpS}(\tokT) 
     \\[4pt]
     \lpFfuni := \lpFfun - \valVi:\uTok{\LpS}(\tokT) 
     \qquad
     \lpMfuni := \lpMfun\setenum{\bind{\uTok{\LpS}(\tokT)}{(\tokT,\valVii-\valV)}}
     \;\; \text{where } (\tokT,\valVii) := \lpMfun(\uTok{\LpS}(\tokT))
   \end{array}}
 {  \tokBal \mid \LpS \mid
    \exchO
    \xrightarrow{\actRedeem{\pmvA}{\valV:\tokT}}
    \tokBal \setenum{\bind{\pmvA}{\tokBali[\pmvA]}} \mid
    (\lpFfuni,\, \lpBfun,\, \lpMfuni) \mid
    \exchO
  }
  \;\nrule{[Rdm]}
\end{equation*}

We exemplify $\redeemOp$ transactions in \Cref{table: example LP redeem}.
From \Cref{table: example LP repay}, $\pmvB$ has a non-zero loan amount, 
hence he can only redeem \mbox{$11:\tokTi[2]$} 
before his collateralization decreases to $\cMin=1.5$, 
at which $\pmvB$ cannot further redeem.
Since $\pmvA$ has no loans, she can redeem as many tokens $\tokTi[0]$ 
as the LP balance permits. 
For $\pmvA$'s redeeming of $50:\tokTi[0]$ for $51:\tokT[0]$ 
the exchange rate is $> 1$, because of
the accrued interest during the prior execution of $\accrueIntOp$. 
% in~\Cref{table: example LP interest}. 
By contrast, the exchange rate for $\pmvB$ is $1$, 
as no loan exists on $\tokTi[2]$, and thus no interest was accrued. 
The tokens $\tokTi[2]$ and $\tokTi[0]$ returned to the LP 
by $\pmvB$ and $\pmvA$ are burnt and subtracted from $\lpMfun$.

% LP-table-redeem
\begin{table}[h!]
\centering
\caption{Running example: redeem actions}
\label{table: example LP redeem}
\scriptsize
\begingroup
\begin{tabular}{|l|c|c|c|c|c|c|c|c|c|c|c|c|l|l|l|c|c|}
\hline
\multirow{2}{*}{Actions} & \multicolumn{4}{c|}{$\tokBal[\pmvA]$} & \multicolumn{5}{c|}{$\tokBal[\pmvB]$} & \multicolumn{3}{c|}{$\lpFfun$} & \multicolumn{3}{c|}{$\lpMfun$} & \multicolumn{2}{c|}{$\Coll{\confG}$} \\ \cline{2-18} 
 & $\tokT[0]$ & $\tokT[1]$ & $\tokTi[0]$ & $\tokTi[1]$ & $\tokT[0]$ & $\tokT[1]$ & $\tokT[2]$ & $\tokTi[0]$ & $\tokTi[2]$ & $\tokT[0]$ & $\tokT[1]$ & $\tokT[2]$ & \multicolumn{1}{c|}{$\tokT[0]$} & \multicolumn{1}{c|}{$\tokT[1]$} & \multicolumn{1}{c|}{$\tokT[2]$} & $\pmvB$ & $\pmvC$ \\ \hline
12. $\actRepay{\pmvC}{15:\tokT[0]}$ & 0 & 150 & 100 & 150 & 0 & 50 & 0 & 50 & 50 & \textbf{135} & 70 & 150 & $\tokTi[0]$:150 & $\tokTi[1]$:150 & $\tokTi[2]$:150 & 1.7 & \textbf{1.9} \\ \hline
13. $\actRedeem{\pmvB}{11:\tokTi[2]}$ & 0 & 150 & 100 & 150 & 0 & 50 & \textbf{11} & 50 & \textbf{39} & 135 & 70 & \textbf{139} & $\tokTi[0]$:150 & $\tokTi[1]$:150 & \textbf{$\tokTi[2]$:139} & \textbf{1.5} & 1.9 \\ \hline
14. $\actRedeem{\pmvA}{50:\tokTi[0]}$ & \textbf{51} & 150 & \textbf{50} & 150 & 0 & 50 & 11 & 50 & 39 & \textbf{84} & 70 & 139 & \textbf{$\tokTi[0]$:100} & $\tokTi[1]$:150 & $\tokTi[2]$:139 & 1.5 & 1.9 \\ \hline
\end{tabular}
\endgroup
\end{table}

\mypar{Liquidation}

When the collateralization of a user $\pmvB$ is 
below the threshold $\cMin$ \circled{6}, another user $\pmvA$ can 
\emph{liquidate} part of $\pmvB$'s loan \circled{8}, in return for a discounted amount of 
minted tokens \emph{seized} from $\pmvB$ \circled{10}. 
$\pmvA$ can execute $\liquidateOp$ if it has enough 
balance to repay a fraction of the lent token \circled{1}, 
and if $\pmvB$ has a sufficient balance of seizable, minted tokens \circled{4}.
The maximum seizable amount % in a liquidation
is bounded by \circled{4} or the resulting collateralization of $\pmvB$ \circled{7},
which cannot exceed $\cMin$. 
After this threshold, $\pmvB$'s collateralization is restored, 
and $\pmvB$ is no longer liquidatable.
\begin{equation*} % \label{eq: LP liquidate transaction}
  \small
  \irule
  {\begin{array}{lll}
     \circled{1}\; \tokBal[\pmvA](\tokT) \geq \valV 
     & \circled{2}\; (\lpBfun\,\pmvB)\,\tokT \geq \valV
     & \circled{3}\; \tokTi \in \minted[\LpS] \\
     \circled{4}\; \tokBal[\pmvB](\tokTi) \geq \valVi
     & \circled{5}\; \valVi = \valV \cdot \frac{\exchO(\tokT)}{\exchO(\uTok{\LpS}(\tokTi))} \cdot \rLiq  \\
     \circled{6}\; \Coll{\tokBal \mid \LpS \mid \exchO}(\pmvB) < \cMin 
     & \circled{7}\; \Coll{\tokBali \mid \LpSi \mid \exchO}(\pmvB) \leq \cMin \\
     \circled{8}\; \lpBfuni := \lpBfun\,\pmvB - \valV:\tokT
     & \circled{9}\; \tokBali[\pmvA] := \tokBal[\pmvA] - \valV:\tokT + \valVi:\tokTi
     & \circled{10}\; \tokBali[\pmvB] := \tokBal[\pmvB] - \valVi:\tokTi \\
   \end{array}
 }
 {\begin{array}{l}
    \tokBal
    \mid \LpS \mid
    \exchO
    \xrightarrow{\actLiquidate{\pmvA}{\pmvB}{\valV:\tokT}{\valVi:\tokTi}}
    \tokBal\setenum{\bind{\pmvA}{\tokBali[\pmvA]}}
    \setenum{\bind{\pmvB}{\tokBali[\pmvB]}} \mid
    (\lpFfun,\lpBfuni,\lpMfun) \mid
    \exchO
  \end{array}}
  \;\nrule{[Liq]}
\end{equation*}

For the execution of $\liquidateOp$, where $\valV:\tokT$ and 
$\valVi:\tokTi$ are repaid and seized amounts respectively, 
the constraint on $\valV$ and $\valVi$ is given in \circled{5}, where: 
\begin{equation} \label{eq: LP liquidation discount}
    \cMin > \rLiq > 1
\end{equation}
The constraint $\rLiq>1$ implies a discount applied to the seized amount 
received by the liquidator, as more value is received than repaid:

% However, $\rLiq$ must be set below the 
% liquidation threshold $\cMin$ to ensure that an undercollateralized 
% account, where $\cMin >\Coll{\tokBal,\LpS,\exchO}(\pmvA) > \rLiq$, 
% can be \emph{recovered} \albertonote{``(i.e. ...'' intuition of what recovered means}. This is illustrated in our example below. 

For the liquidations in \Cref{table: example LP liquidation}, 
we set $\rLiq=1.1$. 
After the price update in action 15, 
both $\pmvB$ and $\pmvC$ are undercollateralized. 
$\pmvC$ is liquidated by $\pmvA$ in transaction 16, 
which restores $\Coll{\confG}(\pmvC)$ to 1.5. 
By contrast, $\Coll{\confG}(\pmvB)$ is 0.9 after the price update. 
Subsequent liquidations by $\pmvA$ seize units of both 
$\tokTi[0]$ and $\tokTi[2]$ until $\pmvB$'s balance 
of minted tokens is empty. 
However, $\pmvB$ still has a loan amount of $11:\tokT[1]$, which is
\emph{unrecoverable}. 
Both $\pmvB$ and potential liquidators have no 
incentive to repay or liquidate given the lack of collateral. 
% LP-table-liquidation
\begin{table}[h!]
\caption{Running example: liquidation actions}
\label{table: example LP liquidation}
\resizebox{\textwidth}{!}{
\scriptsize
\begingroup
\begin{tabular}{|l|c|c|c|c|c|c|c|c|c|c|c|c|c|c|c|c|c|c|c|c|c|c|c|c|}
\hline
\multirow{3}{*}{Actions} & \multicolumn{5}{c|}{\multirow{2}{*}{$\tokBal[\pmvA]$}} & \multicolumn{5}{c|}{\multirow{2}{*}{$\tokBal[\pmvB]$}} & \multicolumn{4}{c|}{\multirow{2}{*}{$\tokBal[\pmvC]$}} & \multicolumn{3}{c|}{\multirow{2}{*}{$\lpFfun$}} & \multicolumn{3}{c|}{$\lpBfun$} & \multicolumn{2}{c|}{\multirow{2}{*}{$\exchO$}} & \multicolumn{2}{c|}{\multirow{2}{*}{$\Coll{\confG}$}} \\ \cline{19-21}
 & \multicolumn{5}{c|}{} & \multicolumn{5}{c|}{} & \multicolumn{4}{c|}{} & \multicolumn{3}{c|}{} & $\pmvB$ & \multicolumn{2}{c|}{$\pmvC$} & \multicolumn{2}{c|}{} & \multicolumn{2}{c|}{} \\ \cline{2-25} 
 & $\tokT[0]$ & $\tokT[1]$ & $\tokTi[0]$ & $\tokTi[1]$ & $\tokTi[2]$ & $\tokT[0]$ & \multicolumn{1}{l|}{$\tokT[1]$} & $\tokT[2]$ & $\tokTi[0]$ & $\tokTi[2]$ & \multicolumn{1}{l|}{$\tokT[0]$} & $\tokT[1]$ & $\tokT[2]$ & $\tokTi[0]$ & $\tokT[0]$ & $\tokT[1]$ & $\tokT[2]$ & $\tokT[1]$ & $\tokT[0]$ & $\tokT[1]$ & $\tokT[0], \tokT[1]$ & $\tokT[2]$ & $\pmvB$ & $\pmvC$ \\ \hline
15. $\exchUpdateOp$ & 51 & 150 & 50 & 150 & - & 0 & 50 & 11 & 50 & 39 & 15 & 30 & 0 & 100 & 84 & 70 & 138 & 59 & 18 & 36 & 1 & \textbf{1.7} & \textbf{0.9} & \textbf{1.2} \\ \hline
16. $\actLiquidate{\pmvA}{\pmvC}{27:\tokT[0]}{50:\tokTi[2]}$ & 51 & \textbf{123} & 50 & 150 & \textbf{50} & 0 & 50 & 11 & 50 & 39 & 15 & 30 & 0 & \textbf{50} & 84 & \textbf{97} & 138 & 59 & 18 & \textbf{9} & 1 & 1.7 & 0.9 & \textbf{1.5} \\ \hline
17. $\actLiquidate{\pmvA}{\pmvB}{27:\tokT[0]}{50:\tokT[0]}$ & 51 & \textbf{96} & \textbf{100} & 150 & 50 & 0 & 50 & 11 & \textbf{0} & 39 & 15 & 30 & 0 & 50 & 84 & \textbf{124} & 138 & \textbf{32} & 18 & 9 & 1 & 1.7 & \textbf{0.7} & 1.5 \\ \hline
18. $\actLiquidate{\pmvA}{\pmvB}{21:\tokT[0]}{39:\tokTi[2]}$ & 51 & \textbf{75} & 100 & 150 & \textbf{89} & 0 & 50 & 11 & 0 & \textbf{0} & 15 & 30 & 0 & 50 & 84 & \textbf{145} & 138 & \textbf{11} & 18 & 9 & 1 & 1.7 & \textbf{0} & 1.5 \\ \hline
\end{tabular}
\endgroup}
\end{table}

\mypar{Transfer of minted tokens} 

Minted tokens can be transferred between users.
Unlike free tokens transfers (rule~\nrule{[Trf]} at page~\pageref{eq:Trf}), 
this requires that the sender retains 
a collateralization level above $\cMin$. 
% LP-minted tokens are not free \circled{3}: 
% $\minted[\LpS] \cap \TokUF = \emptyset$.
\begin{equation*} \label{eq: LP token transfer rule}
  \small
  \irule
  {\tokBal[\pmvA](\tokT) \geq \valV 
   \qquad
   \tokT \in \minted[\LpS] 
   \qquad
   \tokBali =
   \tokBal
   \setenum{\bind{\pmvA}{\tokBal[\pmvA] - \valV:\tokT}}
   \setenum{\bind{\pmvB}{\tokBal[\pmvB] + \valV:\tokT}} 
   \qquad
   \Coll{\tokBali \mid \LpS \mid \exchO}(\pmvA) \geq \cMin 
 }
 {
   \tokBal \mid \LpS \mid \exchO
   \xrightarrow{\actTransferM{\pmvA}{\pmvB}{\valV:\tokT}}
   \tokBali \mid \LpS \mid \exchO
 }
 \;\nrule{[Mtrf]}
\end{equation*}

\mypar{Price updates} 

Finally, the price oracle can be updated non-deterministically:
\begin{equation*} \label{eq: token price update}
  \small
  % \irule{}
  {
    \tokBal \mid \LpS \mid \exchO
    \xrightarrow{\exchUpdateOp} 
    \tokBal \mid \LpS \mid \exchOi
  }
  \;\;\;\nrule{[Px]}
\end{equation*}

%% file: lp-properties.tex
\section{Fundamental properties of lending pools}
\label{sec:lp-properties}

We now establish some fundamental properties of lending pools.
These properties hold for all \emph{reachable states},
\ie states $\confG$ such that $\confG[0] \xrightarrow{}^* \confG$
for some initial $\confG[0]$.

%\Cref{lma: LP only loans in free tokens} states that
%users can only borrow free tokens.
%As such, they cannot be seized by a liquidation,
%and they are freely transferable between users.

%\begin{lemma}  \label{lma: LP only loans in free tokens}
%  Let $\confG$ be a reachable state.
%  % $\pmvA$ be an agent, $\valV$ be a natural number and $\tokT$ be a token. If the following transaction can be executed
%  If
%  \(
%  \confG \xrightarrow{\actBorrow{\pmvA}{\valV:\tokT}} \confGi
%  \),
%  then $\tokT \in \TokUF$.
%\end{lemma}
%\begin{proof}
%  % By inspection of rules \nrule{[Bor]} and \nrule{[Dep]}
%  % in~\Cref{eq: LP borrow transaction}.
%  Premise \circled{1} in rule \nrule{[Bor]}
%  requires that the borrowed token $\tokT$
%  exists in the domain of the component $\lpFfun$ of the state.
%  The only rule which extends $\dom{\lpFfun}$ is~\nrule{[Dep]},
%  where the premise \circled{5} requires $\tokT \in \TokUF$.
%\end{proof}

The first property states that the component $\lpMfun$ of the state
correctly records the balance of all minted tokens held by users.
This is formalized by~\Cref{lma: LP consistent minted token supply}.

% This implies that the supply of LP-minted tokens is tracked correctly
% in our model.

\begin{lemma} \label{lma: LP consistent minted token supply}
  Let $\tokBal \mid \LpS \mid \exchO$ be a reachable state.
  % where $\minted[\LpS] \neq \emptyset$, bart: not needed
  For all $\tokT \in \minted[\LpS]$:
  \begin{equation} \label{eq: LP consistent minted token supply}
    \textstyle
    \sum_{\pmvA} \tokBal[\pmvA](\tokT)
    \; = \;
    snd(\lpMfun(\uTok{\LpS}(\tokT)))
  \end{equation}
\end{lemma}
Another crucial property is that
the exchange rate of a minted token must either strictly increase, 
when users are borrowing the underlying token, or remain 
stable otherwise. This guarantees a depositor that her 
deposit will grow.

\begin{lemma}
  \label{lma:ER-increasing}
  Let \mbox{$\tokBal \mid \LpS \mid \exchO$} be a reachable state, let
  \mbox{$\tokBal \mid \LpS \mid \exchO \xrightarrow{\ltsLabel} \tokBali \mid \LpSi \mid \exchOi$}, 
  and let $\tokT \in \minted[\LpS]$. 
  Then:
  \begin{inlinelist}[(a)]
  \item if $\ltsLabel = \accrueIntOp$ and 
    $\exists\pmvA : (\lpBfun\,\pmvA)\,\tokT > 0$,
    then $\ER[\LpS]{\tokT} < \ER[\LpSi]{\tokT}$;
  \item otherwise, 
    $\ER[\LpS]{\tokT} = \ER[\LpSi]{\tokT}$.
  \end{inlinelist}
\end{lemma}

As a direct consequence of~\Cref{lma:ER-increasing} we have that, 
in any computation, the exchange rate of any token type is increasing.
% Note that from the proof of Lemma~\ref{lma:ER-increasing}
% it also follows that the exchange rate of minted tokens will \emph{strictly}
% increase upon execution of interest accrue.

We also establish a preservation property of the
\emph{supply} of any free token~$\tokT$,
\ie the sum of all user balances and lending balance of $\tokT$:
\begin{equation}
\label{eq: free token supply}
    \supply{\confG}(\tokT)
    \; = \;
    \lpFfun(\tokT) +
    \textstyle \sum_{\pmvA} \tokBal[\pmvA](\tokT)
    \qquad \text{if $\confG = \tokBal \mid \LpS \mid \exchO$}
\end{equation}
\Cref{lma:constant-free-token-supply}
establishes that the supply of any free token is constant. 
% in all reachable states.

\begin{lemma}
  \label{lma:constant-free-token-supply}
  Let $\confG[0] \xrightarrow{}^* \confG$, for $\confG[0]$ initial.
  For all $\tokT \in \TokUF$:
  $\supply{\confG[0]}(\tokT) = \supply{\confG}(\tokT)$.
 \end{lemma}

While the supply of an LP remains constant, 
users act to increase their own share.    
We define the \emph{net worth} 
$W_\Gamma(\pmvA)$ as the value 
of the amount of tokens in $\pmvA$'s wallet or lent by $\pmvA$, 
minus the value of $\pmvA$'s loans.
Formally, if $\confG = \tokBal \mid \LpS \mid \exchO$:
\[
W_\Gamma(\pmvA) = 
\textstyle \sum_{\tokT \in \TokUF}  
\big( 
\tokBal[\pmvA](\tokT) 
+ \tokBal[\pmvA] (\fst(\lpMfun(\tokT)) \cdot \ER[\LpS]{\tokT} 
- (\lpBfun \, \pmvA \, \tokT)
\big) 
\cdot \exchO(\tokT)
%
% \sum_{\tokT \in \TokUF} (\lpBfun \, \pmvA) \tokT \cdot \exchO(\tokT)
\]
The net worth of a user can be increased in short or long sequences of transitions. In general, there is no winning strategy (in the game-theoretic sense) for a single user that wants to increase her net worth, unless she can control price updates: 
actually, with just one price update the net worth of any user 
can be reduced to 0.  
However, under certain conditions, winning strategies can be found. 
We consider first a simple $1$-player game where a user can choose her next action to improve her net worth in the next state. 
Here, liquidation is the only action by an honest user $\pmvA$ that increases her net worth in just one transition. 
\begin{lemma}\label{lma:game0}
%
%Let $\Gamma$ be a reachable state and $\Gamma \xrightarrow{\ell} \Gamma'$ such that $\ell$ is none of $\accrueIntOp$, $\actLiquidate{\pmvA}{\pmvB}{\valV:\tokT}{\valVi:\tokTi}$, $\actTransferM{\pmvA}{\pmvB}{\valV:\tokT}$, $\actTransfer{\pmvA}{\pmvB}{\valV:\tokT}$ and $\exchUpdateOp$. Then for all $\pmvA \in \PmvU$ it holds $W_\Gamma(\pmvA) = W_{\Gamma'}(\pmvA)$.
%
Let $\Gamma$ be a reachable state and $\Gamma \xrightarrow{\ell} \Gamma'$ 
with $\ell = {\_}_{\pmvA}(\cdots)$. 
Then: 
\begin{inlinelist}[(a)] 
\item $W_\Gamma(\pmvA) > W_{\Gamma'}(\pmvA)$ 
% if $\ell = \actLiquidate{\pmvA}{\pmvB}{\valV:\tokT}{\valVi:\tokTi}$;
if $\ell = \liquidateOp_{\pmvA}(\cdots)$;
\item $W_\Gamma(\pmvA) = W_{\Gamma'}(\pmvA)$ otherwise.
\end{inlinelist}
\end{lemma}
Since this is the winning strategy for all users, but liquidations may be limited by loan or collateral amounts, an adversary who has the power to drop or reorder transactions can potentially monopolize liquidations for itself. 
We refer to \Cref{sec:related} for additional discussion of such attacks.

We now consider a slightly extended game, 
where $\pmvA$ guesses that the next (adversarial) action is going to be $\ell$, resulting in $\Gamma_0\xrightarrow{\ell} \Gamma_1$ but can still perform an action $\ell'$ before $\ell$, resulting in $\Gamma_0 \xrightarrow{\ell'} \Gamma'_0 \xrightarrow{\ell}  \Gamma'_1$. 
The goal of $\pmvA$ is to choose $\ell'$  such that $W_{\Gamma'_1}(\pmvA) > W_{\Gamma_1}(\pmvA)$.
We show that if $\ell = \accrueIntOp$, 
\ie $\pmvA$ is expecting interest accrual to happen next, 
her choice is limited to deposits, repays and liquidations.

\begin{lemma}\label{lma:game1}
Let $\Gamma_0$ be a reachable state, and let
$\Gamma_0\xrightarrow{\ell} \Gamma_1$ and 
$\Gamma_0 \xrightarrow{\ell'} \Gamma'_0 \xrightarrow{\ell}  \Gamma'_1$ 
be such that $\ell = \accrueIntOp$ and 
$\ell' = {\_}_{\pmvA}(\cdots)$. 
Then: 
\begin{inlinelist}[(a)]
\item $W_{\Gamma'_1}(\pmvA) \ge W_{\Gamma_1}(\pmvA)$ 
if $\ell'$ is one of 
$\liquidateOp_{\pmvA}(\cdots)$ or
% $\actLiquidate{\pmvA}{\pmvB}{\valV:\tokT}{\valVi:\tokTi}$,
$\depositOp_{\pmvA}(\cdots)$ or
% $\actDeposit{\pmvA}{\valV:\tokT}$ 
$\repayOp_{\pmvA}(\cdots)$; 
% $\actRepay{\pmvA}{\valV:\tokT}$; 
\item $W_{\Gamma'_1}(\pmvA) \leq W_{\Gamma_1}(\pmvA)$ otherwise. 
\end{inlinelist}
\end{lemma}

Overall, \Cref{lma:game0,lma:game1} 
determine the set of  actions to consider (together with their parameter) to maximize improvements in short-term net worth. 

% \begin{lemma}
% \[
% \begin{array}{ll}
%     \ER[\LpS]{\tokT} < \ER[\LpSi]{\tokT} \quad\;
%     & \exists\pmvA\;s.t.\;(\lpBfun\,\pmvA)\,\tokT > 0 \\
%     \ER[\LpS]{\tokT} = \ER[\LpSi]{\tokT}
%     & \mathit{otherwise}
% \end{array}
% \]
% \end{lemma}

%% file: lp-attacks.tex
\section{Lending pool safety, vulnerabilities and attacks}
\label{sec:lp-attacks}

We discuss further properties of lending pools, 
focusing on potential risks which could lead to unsecured loans or exploitations by malicious actors. 
In particular, we focus on user collateralization and the availability of free token funds in lending pools (utilization). In the case where these can be targeted by an attacker, the motivation is to limit the lending pool functionality (denial-of-service) or cause the victim to incur losses, which in some cases may imply a gain for the attacker.
We restrict our attention to attacker models where the attacker has the ability to perform some of the actions of the LP model, or even update the price oracle. More powerful attackers that can drop or reorder transactions are discussed in~\Cref{sec:related}.

\subsection{Collateralization bounds and risks}
The lending pool design assumes that loans are \emph{secured} by collateral: liquidations thereof are incentivised in order to recover loans should the borrowing users fail to repay. However, collateral liquidation is exposed to risks. Firstly, the incentive to liquidate is only effective, if the liquidator values the seized collateral higher than the value of the repaid loan amount, implying a profit. Secondly, large fluctuations in token price may reduce the relative value of the collateral such that the loan becomes partially unrecoverable. Furthermore, an attacker with the ability to update token prices can force users to become undercollateralized and then seize the collateral of victims without repaying any loans.

\mypar{LP-minted token risk}
The lending pool must determine the appropriate levels of collateralization based on token prices given by the price oracle. However, the value of LP-minted tokens is indeterminable since they are not featured in in $\dom(\exchO)$. The definition of collateralization in \cref{eq: LP account collateralization} values units of LP-minted tokens at the same price as their underlying counterpart, as do lending pool implementations~\cite{aaveatokpr,compctokpr}. However, since LP-minted tokens represent claims on
free tokens, which are only redeemable if sufficient funds remain in the
lending pool (\circled{2} in \nrule{[RDM]}), it is possible that users value
minted tokens at a lower price than their underlying counterparts when LP-minted tokens cannot be redeemed during times of low lending pool funds (utilization). 
Lending pool designs do not account for this and thus run the risk of incorrectly pricing LP-minted tokens and collateral.

\mypar{Safe collateralization}
Assuming a correct valuation of LP-minted tokens, undercollateralized loans should be swiftly liquidated, given the incentivization provided by the liquidation discount. Furthermore, the user collateral value should be high enough, such that the user's loan amount is sufficiently repaid by liquidations to recover the user collateralization back to $\cMin$. Therefore, we introduce two notions of safe collateralization.

Inspired by \cite{kaoanalysis}, we say that a LP state is \emph{$\epsilon$-collateralization safe}
when the ratio of the loan value of undercollateralized accounts 
to the total loan value of the lending pool
is below the threshold $\epsilon$:
\begin{equation}
  \label{eq:fracUcLoanVal}
  % \fracUcLoanVal{\tokBal,\LpS,\exchO} =
  \frac{
    \sum_{\Coll{\confG}(\pmvA) < \cMin} \loanVal[\confG](\pmvA)
  }{
    \sum_{\pmvA} \loanVal[\confG](\pmvA)
  } 
  \leq 
  \epsilon
\end{equation}
If the liquidation incentive is effective, 
% a $\fracUcLoanVal{\tokBal,\LpS,\exchO}$ 
a value below $\epsilon$ should not persist, 
as users are quick to execute liquidations. 
The efficiency of lending pool liquidations 
has been studied in \cite{perez2020liquidations}. 
We note that sufficiently large volumes of seized 
collateral which are immediately sold on external markets
may delay further liquidations,
as investigated in \cite{gudgeon2020decentralized},
due to the external market's finite capacity to
absorb such a sell-off.

However, \emph{$\epsilon$-collateralization safety} does not account for undercollateralized loans 
which are \emph{non-recoverable}, as previously illustrated in the example of 
\Cref{table: example LP liquidation}. The set of 
non-recoverable, undercollateralized accounts are those with a 
collateralization below $\rLiq$.
The non-recoverable loan value of an account is given by $\nrLoanVal{\confG}$. 
It represents the remaining loan value 
of a user $\pmvA$ should it be fully liquidated, 
such that no further collateral can be seized.
\begin{equation} \label{eq: LP unrecoverable loan value}
  \nrLoanVal{\confG}(\pmvA) = \begin{cases}
    \loanVal[\confG](\pmvA)
    -
    \frac{
      \collVal[\confG](\pmvA)
    }{\rLiq}
    & \mathit{iff} \;\Coll{\confG}(\pmvA) < \rLiq
    \\
    0
    & \mathit{otherwise} 
  \end{cases}
\end{equation}
\Cref{eq: LP unrecoverable loan value} illustrates that for the case where
an account collateralization is below $\rLiq$, the discounted value of the 
collateral can no longer equal or exceed the remaining loan value,
a consequence of \eqref{eq: LP account collateralization} and 
\eqref{eq: LP liquidation discount}.
We say that a LP state is \emph{strongly $\epsilon$-collateralization safe}
when the fraction of the total loan value of a lending pool 
which is not recoverable is below $\epsilon$:
\begin{equation} \label{eq: LP fraction of non-recoverable loans}
  % \fracNrUcLoanVal{\tokBal,\LpS,\exchO} =
  \frac
  {
    \sum_{\pmvA}
    \nrLoanVal{\confG}(\pmvA)
  }
  {
    \sum_{\pmvA}
    \loanVal[\confG](\pmvA)
  }
  \leq
  \epsilon
\end{equation}

The condition~\eqref{eq: LP fraction of non-recoverable loans} is
actually stronger than~\eqref{eq:fracUcLoanVal}, \ie 
if a state is strongly $\epsilon$-collateralization safe,
then it is also $\epsilon$-collateralization safe. Given equal 
denominators of~\eqref{eq:fracUcLoanVal} 
and~\eqref{eq: LP fraction of non-recoverable loans}, 
this is a consequence of comparing numerators:
Here, it can be observed that the numerator of~\eqref{eq:fracUcLoanVal} 
is greater than that of~\eqref{eq: LP fraction of non-recoverable loans}, 
as $\loanVal[\confG](\pmvA)$ is necessarily greater than 
$\nrLoanVal{\confG}(\pmvA)$ by definition and 
the set $\left\{ \pmvA \mid \Coll{\confG}(\pmvA) < \cMin \right\}$
is a superset of 
$\left\{\pmvA \mid \Coll{\confG}(\pmvA) < \rLiq \right\}$~\eqref{eq: LP unrecoverable loan value}.

Strong price volatility is a risk to $\epsilon$-collateralization safety, as a sharp drop in price can immediately reduce a previously sufficiently collateralized user to become undercollateralized below the threshold of $\cMin$: such an immediate drop leaves the user with no opportunity maintain its collateral with repayments.

\mypar{Attacks on safe collateralization} 
Malicious agents which can perform price updates can therefore influence the evolution of the LP to lead it to a state that is not $\epsilon$-collateralization safe or strongly $\epsilon$-collateralization safe. 

%\mypar{Price oracle attack} 

%Thus far, we have assumed that the price 
%oracle $\exchO$ in a reachable state is updated non-deterministically. 
% We note that our model can be
% extended to model a price oracle which can be arbitrarily updated by an 
% attacker from a set of dishonest users $\PmvU_m$.
% \begin{equation} \label{eq: LP oracle manipulation}
%     \irule
%     {\begin{array}{c}
%         \pmvA \in \PmvU_m
%     \end{array}}
%     {
%         \tokBal \mid \LpS \mid \exchO 
%         \xrightarrow{\actExchUpdate{\pmvA}{\valV:\tokT}}
%         \tokBal \mid \LpS \mid \exchO\setenum{\bind{\tokT}{\valV}} 
%     }
% \end{equation}
% \noindent
For example, a malicious agent controlling the price oracle could act as follows. First, she would perform price updates to push any account 
collateralization below $\cMin$, such that it becomes undercollateralized.
The attacker can then perform liquidations on these accounts and
benefit from the discount resulting from both the price update and $\rLiq$. 
The attacker has maximized her profits by updating $\exchO$ such that 
$\loanVal[\confG](\pmvB)$ in~\eqref{eq: LP account collateralization} is zero, 
where $\pmvB$ is an account under attack. 
In this case, $\actLiquidate{\pmvA}{\pmvB}{\valV:\tokT}{\valVi:\tokTi}$ 
can be performed with $\valV = 0$, and repeated liquidations 
can be executed to seize the full balance of $\pmvB$'s LP-tokens. 

As a matter of fact, a recent failure of the oracle price feed utilized by the Compound lending pool implementation lead to \$100M of collateral being (incorrectly) liquidated \cite{compoundoracleattack}: though it is unclear whether this
was an intentional exploit, it illustrates the feasibility of such a
price oracle attack.

\subsection{Utilization bounds and risks} 
The notion of \emph{utilization} plays a fundamental role in the incentive model of lending pools as explained in~\cite{gudgeon2020plf}. As a matter of fact, it is often used as a key parameter of interest rate models in implementations \cite{aave,comp} and literature \cite{gudgeon2020plf}.
The utilization of a token type in a lending pool
is the fraction of previously deposited funds currently
lent to borrowing users. 
Formally:
\begin{equation} \label{eq: LP pool utilization}
    \Util{\LpS}(\tokT) 
    \; = \; 
    \dfrac{\sum_{\pmvA} (\lpBfun \, \pmvA) \, \tokT}
    {\lpFfun(\tokT) + \sum_{\pmvA} (\lpBfun \, \pmvA) \, \tokT}
    %\qquad \text{if $\lpFfun(\tokT) > 0$ or $\exists \pmvA : (\lpBfun \, \pmvA) \, \tokT > 0$}
\end{equation}

\mypar{Over- and under-utilization}
The value of $\Util{\LpS}(\tokT)$ ranges between 0 and 1.
We say that $\tokT$ is \emph{under-utilized} if its utilization is 0 and \emph{over-utilized} when it is 1. We say that an LP state is under(over)-utilized if there is at least one under(over)-utilized token. 

Under-utilization occurs when some units of $\tokT$ have been deposited, 
but not lent to any user.
This implies that action $\accrueIntOp$ 
%in  \cref{eq: LP interest compounding} 
does not increase the loan value of 
any account, so that the exchange rate of $\tokT$ in 
\eqref{eq: LP exchange rate} remains constant, thereby not resulting in any gain for lenders.  

On the other hand, over-utilization occurs when some users have borrowed $\tokT$,
but the lending pool has no deposited funds of $\tokT$. In this case users can neither borrow or redeem.  

Under- and over-utilization are not desirable and should be avoided.  
An optimal utilization rate 
\cref{eq: LP pool utilization} of a free token type $\tokT$ strikes a
balance between the competing objectives of interest maximization and 
the ability for users to borrow or redeem tokens of type $\tokTi = \fst(\lpMfun(\tokT))$. 
In particular, the lending pool interest rate models described in~\cite{gudgeon2020plf}
intend to incentivize actions of both borrowers and lenders to discover a utilization
equilibrium between under- and over-utilization. Informally, this is achieved with interest rate models 
which rise and fall with utilization: 
increasing utilization and interest rates incentivize deposits and
repayment of loans. Decreasing utilization and interest rates incentivize redeems
and additional loan borrowing. 

We proceed to discuss under- and overutilization attacks: here, we note that the former is weaker than the latter, as funds can still be safely recovered in a case of underutilization.

\mypar{Under-utilization attacks}
% Attacks on safe utilization towards 0
Under-utilization can be achieved by a group of malicious users interested in reducing interest accrual for depositors or discouraging borrowing of a token $\tokT$. Here, the attacker can temporarily reduce utilization by repaying large amounts of loans, though the effectiveness of this approach will depend on the amounts of $\tokT$ repaid by the attacker, as a lowered utilization can also reduce the interest rate (in certain models \cite{gudgeon2020plf}), thereby incentivizing additional borrowing. An attacker which can update the price oracle can lower the collateralization of borrowers arbitrarily, thereby incentivizing repayments and liquidations to target lower utilization of specific tokens.

\mypar{Over-utilization attacks}
Over-utilization could be achieved by a group of malicious users interested in preventing redeems or borrows of $\tokT$. 
The malicious users can do this by redeeming all units of $\tokT$ while avoiding loans to be repaid or liquidated.
We illustrate an over-utilization attack in~\Cref{tab:lending-pool:over-util-attack}. Here, users $\pmvA$ and $\pmvC$ initially hold the entire supply of $\tokT[0]$ in their balances. $\pmvA$ colludes with $\pmvB$ to steal $\pmvC$'s balance of $\tokT[0]$: in actions 0-2, both $\pmvA$ and $\pmvB$ deposit units of $100:\tokT[0]$ and $100:\tokT[1]$ respectively. $\pmvB$ utilizes her balance of $100:\tokTi[1]$ as collateral to borrow $50:\tokT[0]$ from the lending pool in action 3. At this point, $\pmvA$ and $\pmvB$ are acting as lender and borrower of $\tokT[0]$, for which the utilization is 0.5. $\pmvC$, having observed an opportunity to earn interest on $\tokT[0]$ decides to deposit $50:\tokT[0]$ in action 4. However, user $\pmvA$ still has a balance of redeemable $100:\tokTi[0]$, which she redeems in action 5. Now, users $\pmvA$ and $\pmvB$ have removed all units of $\tokT[0]$ from the lending pool, pushing the utilization of $\tokT[0]$ to 1 and preventing $\pmvC$ from redeeming her funds. Of course, user $\pmvB$ cannot redeem his balance of $\tokTi[1]$ since her loan has not been repaid, but this can be considered the cost of the attack.

\begin{table}[t!]
\centering
\scriptsize
\begingroup
\begin{tabular}{|l|c|c|c|c|c|c|c|c|c|c|c|c|c|c|}
\hline
\multirow{2}{*}{Actions} & \multicolumn{2}{c|}{$\tokBal[\pmvA]$} & \multicolumn{3}{c|}{$\tokBal[\pmvA]$} & \multicolumn{2}{c|}{$\tokBal[\pmvC]$} & \multicolumn{2}{c|}{$\lpFfun$} & $\lpBfun \,\pmvB$ & \multicolumn{2}{c|}{$\lpMfun$} & \multicolumn{2}{c|}{$\Util{\LpS}$} \\ \cline{2-15} 
 & $\tokT[0]$ & $\tokTi[0]$ & $\tokT[0]$ & $\tokT[1]$ & $\tokTi[1]$ & $\tokT[0]$ & $\tokTi[0]$ & $\tokT[0]$ & $\tokT[1]$ & $\tokT[0]$ & $\tokT[0]$ & $\tokT[1]$ & $\tokT[0]$ & $\tokT[1]$ \\ \hline
0. Initial State & 100 & -- & -- & 100 & -- & 50 & -- & -- & -- & -- & -- & -- & -- & -- \\ \hline
1. $\actDeposit{\pmvA}{100:\tokT[0]}$ & \textbf{0} & \textbf{100} & -- & 100 & -- & 50 &  & \textbf{100} & -- & -- & \textbf{$\tokTi[0]$:100} & -- & \textbf{0} & -- \\ \hline
2. $\actDeposit{\pmvB}{100:\tokT[1]}$ & 0 & 100 & -- & \textbf{0} & \textbf{100} & 50 &  & 100 & \textbf{100} & -- & $\tokTi[0]$:100 & \textbf{$\tokTi[1]$:100} & 0 & \textbf{0} \\ \hline
3. $\actBorrow{\pmvB}{50:\tokT[0]}$ & 0 & 100 & \textbf{50} & 0 & 100 & 50 &  & 50 & 100 & \textbf{50} & $\tokTi[0]$:100 & $\tokTi[1]$:100 & \textbf{0.5} & 0 \\ \hline
4. $\actDeposit{\pmvC}{50:\tokT[0]}$ & 0 & 100 & 50 & 0 & 100 & \textbf{0} & \textbf{50} & 100 & 100 & 50 & \textbf{$\tokTi[0]$:150} & $\tokTi[1]$:100 & \textbf{0.3} & 0 \\ \hline
5. $\actRedeem{\pmvA}{100:\tokTi[0]}$ & \textbf{100} & \textbf{0} & 50 & 0 & 100 & 0 & 50 & \textbf{0} & 100 & 50 & \textbf{$\tokTi[0]$:50} & $\tokTi[1]$:100 & \textbf{1.0} & 0 \\ \hline
\end{tabular}
\endgroup
\caption{Over-utilization attack.}
\label{tab:lending-pool:over-util-attack}
\end{table}

% Alice and and Bob are the only users that have $\tokT$ in their wallets.  
% Alice is in coalition with Charlie want Bob to loose (part of) his $\tokT$-liquidity. No $\tokT$ has been ever deposited on the LP so Charlie is not incentivized to lend.
% \begin{enumerate}
% \item Alice deposits $x$ $\tokT$. Still no incentive for Charlie. LP is unsafe now (utilization of $\tokT$ is 0) but Alice and Bob do not care.
% \item Bob borrows $y$ $\tokT$ to make the LP safe and to incentivize Charlie to lend.
% \item Charlie deposits $z$ $\tokT$ keeping the LP is still safe.
% \item Alice withdraws $u$ $\tokT$ preventing Charlie to withdraw, and making the LP unsafe so no other user is incentivized to deposit or borrow. 
% \end{enumerate}
% Find suitable $x,y,z,u$ and $\epsilon$s. Bob is ``sacrified'' but I assume Alice and Bob can afford it.

% Attack on safe utilization towards 1

%% file: related.tex
\section{DeFi archetypes: lending pools and beyond}
\label{sec:related}

% There is not a well-accepted definition of the term ``Decentralized Finance'' with precisely formulated inclusion or exclusion criteria. 
% Instead, we put our attention on smart contract archetypes commonly referred to as DeFi applications in documentation or literature. 
% We note that such contracts generally exhibit strong interoperability due to standardized token interfaces \cite{ERC20}.
%
% Furthermore, we emphasize prior work which contributes functional descriptions of DeFi archetype behaviour, rather than purely economic or taxonomic investigations thereof.
%
We now discuss the interplay between lending pools and other DeFi applications,
like algorithmic stable coins, automatic market makers, margin trading and flash loans, which are all predominantly deployed on the Ethereum blockchain \cite{ethereum}. 
% Where applicable, we relate these investigations to lending pools and our proposed model.

\mypar{Lending pools} 

The emergent behaviour of lending pools in times of high price volatility is examined in \cite{gudgeon2020decentralized} by simulation of a lending pool liquidation model. Here, a large price drop can cause many accounts to become undercollateralized: assuming liquidators sell off collateral at an external market for units of the repaid token type, the authors suggest that limited market demand for collateral tokens may prevent liquidations from being executed, thereby posing a risk to \emph{$\varepsilon$-collateralization safety} as we have defined in~\cref{eq:fracUcLoanVal,eq: LP fraction of non-recoverable loans}.

Lending pool behaviour at the user level is modelled in \cite{kaoanalysis}, which simulates agents interacting with the Compound implementation to examine the evolution of \emph{liquidatable} and \emph{undercollateralized} debt, notions similar to \emph{(strong) $\epsilon$-collateralization safety}~\eqref{eq:fracUcLoanVal}~\eqref{eq: LP fraction of non-recoverable loans}. \cite{chitra2019competitive,chitra2020stake} examine the competition for user deposits between staking in proof-of-stake systems and lending pools: in the case where lending pools are believed to be more profitable, users may shift deposits away from the staking contract of the underlying consensus protocol towards lending pools, thereby endangering the security of the system.

Lending pool interest rate behaviour is examined in \cite{gudgeon2020plf}, where empirical behaviour of interest rate models in Compound \cite{comp}, Aave \cite{aave} and dYdX \cite{dydx} are analyzed. In particular, the authors observe a statistically significant coupling in interest rates between deployed lending pools, suggesting that the dynamic interest models are effective in discovering a global interest rate equilibrium for a given token. 
Our formal model is parameterized by the interest rate, 
that must always be positive \eqref{eq: interest rate}: 
since this property holds for all interest rate functions in \cite{gudgeon2020plf}, 
our model can be instantiated with them.

\mypar{Algorithmic stable coins} 
MakerDAO \cite{makerdao} is the leading algorithmic stable coin and is credited with being one of the earliest DeFi projects. It incorporates several features found in lending pools, such as deposits, minting, and collateralization. 
Users are incentivized to interact with the smart contract to mint or redeem DAI tokens. This, in turn, adjusts the supply of DAI such that a stable value against the reference price (e.g USD) is maintained. Synthetic tokens are similar to algorithmic stable coins but may track an asset price such as gold or other real-world assets. Reference asset prices are determined by price oracles.

The authors of \cite{moin2019classification} introduce a taxonomy for various price stabilization mechanisms, providing insight into the functionality of such contracts. \cite{gudgeon2020decentralized} uncovers a vulnerability in the governance design of MakerDAO, allowing an attacker to utilize flash loans to steal funds from the contract. The empirical performance of MakerDAO's oracles is studied in \cite{gu2020empirical}, which also proposes alternate price feed aggregation models to improve oracle accuracy. Finally, \cite{darlin2020optimal} investigates the optimal bidding strategy for collateral liquidators in MakerDAO, which is executed by through user auctions.

Stable coins which track prices of real-world currencies (e.g. USD) exhibit a price stability useful for lending pools: users with stable collateral or loan values have a lower likelihood of suddently becoming undercollateralized.

\mypar{Automatic market makers} 
Leading automatic market makers Uniswap \cite{uniswap} and Curve Finance \cite{curvestats} hold \$1.6B \cite{uniswapstats} and \$1.5B \cite{curvestats} worth of tokens and feature an estimated \$320M \cite{uniswapstats} and \$36M \cite{curvestats} worth of token exchange transactions every day. An automatic market maker (AMM) is organized in token pairs $(\tokT, \tokTi)$, which users can interact with to exchange units of $\tokT$ for $\tokTi$ or vice-versa. AMM's do not match opposing actions of buyers and sellers: users simply exchange tokens with a AMM pair, where the exchange rate is determined algorithmically as a function of the AMM pair balance. Hence, the dynamic exchange rate of an AMM token pair is affected with each user interaction. 

The work in \cite{angeris2019analysis} investigates alternative, algorithmic exchange rate models and defines the user arbitrage problem, where a profit-seeking agent must determine the optimal set of AMM pairs (with differing exchange rates) to interact with: given such arbitrage opportunities will be exploited by rational users, it is expected that exchange rates across AMM's remain consistent. AMM price models can fail: The \emph{constant product} exchange rate model implemented by Uniswap \cite{uniswap} and Curve \cite{curve} is simple, but can theoretically reach a state where the the exchange rate is arbitrarily high. \cite{wang2020automated} proposes bounded exchange rate models to address this. 

\cite{angeris2020improvedoracles} suggests that AMM's track global average token prices effectively. As such, AMM's can inform price oracles: such oracles, however, only update price information with each new block \cite{uniswaporacle} computed from time-weighted price averages of AMM pairs over the past block interval. This increases the cost of manipulating prices of the oracle, as the manipulated price must be sustained over a period of time. We note that lending pool implementations do not rely on oracles which derive prices from AMM states.

AMM's suffer from front-running, where an attacking user observes the victim's announced, yet unconfirmed token exchange transaction, and sequences its own transaction prior to that of the victim. A front-running attack on an AMM user takes advantage of the change in exchange rate resulting from the victim's token exchange, who ends up paying a higher price, as illustrated in \cite{zhou2020high}. Front-running of smart contracts is investigated more generally in \cite{eskandari2019sok}: mitigations such as commit-and-reveal schemes are proposed, which come with an increased cost for user-contract interactions. In the context of AMM's, \cite{daian2019flash} introduces the notion of gas auctions, where adversarial users compete to front-run a given AMM exchange transaction by outbidding each others transaction fee.

We note that similar attacks can be modeled with an attacker that can drop or reorder transactions in our lending pool model. Such an attacker can trivially defer attempts of a borrower to repay a loan: subsequent interest accrual will eventually cause the user to become undercollateralized, so that the attacker can liquidate the victim. Such an attacker can also monopolize all liquidations for herself, preventing other users from executing such an action: \cite{daian2019flash} suggests that miners may be incentivized to perform such attacks due to gain resulting from liquidation discounts.

\mypar{Margin trading} 
An important use case of lending pools are \emph{leveraged} long or short positions initiated by users, also referred to as margin trading. In a leveraged long position of $\tokT$ against $\tokTi$, the user speculates that the price of the former will increase against the price of the latter: a user borrows $\tokTi$ at a lending pool against collateral deposited in $\tokT$, and then exchanges the borrowed units of $\tokTi$ back to $\tokT$ at a token exchange or an AMM. The user will now earn an amplified profit if the price of $\tokT$ appreciates relative to $\tokTi$, since both the borrowed balance and redeemable collateral in $\tokT$ appreciates in value whilst only the loan repayable with $\tokTi$ decreases in value. A leveraged short position simply reverses the token types. Margin trading contracts such as bZx Fulcrum \cite{fulcrum} combine lending and AMM functionalities to offer margin trades through a single smart contract. However, since such margin trading contracts perform large token exchanges at external AMM's, attackers can use such actions to manipulate AMM prices, as shown in \cite{qin2020attacking}. Furthermore, the scope of such attacks is magnified when performed with flash loans. 
%
% \albertonote{When we mention the price attacks in our model we could somehow link them to this kind of things. That is to explain that it is very realistic that a malicious user can manipulate price oracles, since in practice prices could come from other platforms, subject to such attacks.}

\mypar{Flash loans} 
Any smart contract holding balances of tokens can expose flash loan functionality to users: here, a user can borrow and return a loan within a single atomic transaction group. Informally, we describe an atomic transaction group as an a sequence of actions from a single user, which must execute to completion or not execute at all. Atomic transaction groups can be implemented in Ethereum by user-defined smart contracts \cite{aaveflashloan}, but can also be supported explicitly, such as in Algorand \cite{bart2020formalalgorand}. As such, flash loans are guaranteed to be repaid or not executed at all. The work in \cite{wang2020towards} introduces an initial framework to identify flash loan transactions on the Ethereum blockchain for an analysis of their intended use-cases, which include arbitrage transactions, account liquidations (in lending pools or stable coins) and attacks on smart contracts. We note that our model can be easily extended to encompass flash loan semantics.

Flash loans have been utilized in recent attacks in DeFi contracts \cite{qin2020attacking} \cite{valuedefi} \cite{harvestfin} \cite{origindefi} \cite{akryopolisdefi}. The flashloan attack on bZx Fulcrum described in \cite{qin2020attacking} involves sending the borrowed tokens to a margin trading contract, which, in turn, initiates a large token exchange at an external AMM: here, the large amount of exchanged tokens causes a significant shift in dynamic AMM exchange rate, which represents an arbitrage opportunity exploited by the attacker in several execution steps involving other contracts. Flash loans provide attackers with access to very large token values to initiate attacks.

%% file: conclusion.tex
\section{Conclusions}
\label{sec:conclusions}

We have provided a systematization of knowledge on lending pools 
and their role in DeFi, by leveraging a new model 
which enables formal definitions of lending pool 
properties, vulnerabilities, and attacks.
This work represents a first step towards the rigorous analysis of 
DeFi contracts, improving existing literature with a precise 
executable semantics of interactions beween users and LPs.
% Further, we have given an overview of DeFi smart contracts 
% and their similarities, differences and interplay with lending pools.

\mypar{Differences between our model and LP implementations}

We have synthesised our model from informal descriptions in the literature 
and actual implementation and documentation of lending pools 
Compound~\cite{comp} and Aave~\cite{aave}. 
To distill a usable, succinct model we have abstracted 
away some implementation details, that could be incorporated in the model 
at the cost of a more complex presentation. 
We discuss here some of the main abstractions we made. 
% focusing on differences related to the management of token deposits, loans and liquidations. 

The original implementations of Compound and Aave
gave administrators control over the economic parameters of the LP,
\ie $\cMin$, $\rLiq$, and the interest rate function. 
This made administrators of such early versions privileged users, 
who could in principle prevent honest depositors, borrowers and liquidators 
from withdrawing funds. A Compound administrator, for example, can replace application logic which computes collateralization and authorizes supported tokens \cite{compswapcontroller}.  
Later versions of these platforms have introduced \emph{governance tokens} 
(respectively, COMP and AAVE), 
which are allocated to initial investors
or to LP users, who earn units of such tokens upon each interaction.
Governance tokens allow holders to propose, vote for, 
and apply changes in economic parameters, including interest rate functions. 
By contrast, our model assumes that economic parameters are fixed,
and omits governance tokens.

In implementations, adding a new token type to the LP
must be authorized by the governance mechanisms. 
By contrast, in our model any user can add a new token type to the LP 
by just performing the first deposit of tokens of that type. 
Implementations also allow administrators or governance
to assign weights to each token type.
% This allows to set the thresholds $\cMin$ and $\rLiq$ 
% for each specific user.
This is intended to adjust collateralization and liquidation thresholds
$\cMin$ and $\rLiq$ for the predicted price volatility of token types 
present in a user's loan and collateral. 
Further, implementations require users to pay \emph{fees} upon actions.
These fees are accumulated in a reserve controlled by the governance mechanisms 
of the LP, and intended to act as a buffer in case of unforeseen events.
Our model does not feature token-specific weights and fees.

User liquidations in implementations are limited to repay a maximum fraction 
of the loan amount~\cite{compmaxliq,aavemaxliq}. 
However, this implementation constraint can be bypassed 
by a user employing multiple accounts, so we omit it in our model.

Lending pool platforms implement the update of interest accrual 
in a \emph{lazy} fashion: since smart contracts cannot trigger transactions, 
periodic interest accrual would rely on a trusted user to reliably perform 
such actions, introducing a source of corruption. 
Therefore, interest accrual is performed whenever a user performs an action
which requires up-to-date loan amounts. 
Here, the interest rate in implementations is not recomputed for each time period. 
Instead, a single interest rate is applied to the period since the last interest accrual \cite{aaveintr,compintr} in order to reduce the cost of execution, leading to inaccuracies in loan interest.

% We restrict our model to functionality related to lending and borrowing, 
% and omit smart contract \emph{governance} mechanics used by governance token holding majorities to alter contract parameters: Governance of a smart contract allows 
% intervention when the lending pool does not behave as anticipated, but also introduces
% a potential source of corruption. As such, we believe it is paramount to first 
% investigate the behaviour of such applications based on their primary, financial 
% functionality: We focus on expressing this as executable semantics in our 
% model of lending pools, from which properties and behaviour can be inferred. 
% This, in turn, promises to provide insight on where
% governance intervention may indeed be necessary and advance the discipline of safe DeFi contract design.

\mypar{Comparison with other LP models}

There are few models of lending pools in the literature. 
The \emph{liquidation model} of~\cite{gudgeon2020decentralized} 
is meant to simulate interactions between lending pool liquidations 
and token exchange markets in times of high price volatility. 
Unlike in our presented model, 
\cite{gudgeon2020decentralized} performs liquidations in aggregate, 
and it omits individual user actions. 
The interest rate functions of \cite{gudgeon2020plf} 
formalize various interest rate strategies used by LP implementations, 
and can be seen as complementary to our work. 
Indeed, even if we did not incorporate such functions 
directly in our model (for brevity), 
they could be easily included as instances of $\Intr{\LpS}(\tokT)$ 
in rule \nrule{[Int]}. 
The work \cite{perez2020liquidations} introduces a LP state model, 
which is instantiated with historical user transactions observable 
in the Compound implementation deployed on Ethereum. 
The model abstraction facilitates the observation of state effects of each interaction, and investigates the (historical) latency of user liquidations following the undercollateralization of borrowing accounts.
Aforementioned work prioritizes high-level analysis over model fidelity: 
indeed, the lending pool properties and attacks we present are a 
direct consequence of the precision in our lending pool semantics.
\mypar{Future work}

Our model already allows us to formally establish properties of LPs
(\Cref{sec:lp-properties}), 
and to precisely describe potential attacks to LPs as sequences of user actions 
(\Cref{sec:lp-attacks}). 
This paves the way for future automatic analyses of LPs, 
which could exploit \eg, model checking or automated theorem proving tools. 
% under the consideration of specific agent strategies.
Following the same approach we used in~\Cref{sec:lending-pools}, 
we could devise formal models of other DeFi archetypes, 
like \eg stable coins, automatic market makers, and flash loans.
In this perspective, it would be possible to extend the scope of 
analysis techniques to attacks that exploit the interplay between 
different archetypes, 
which so far have been found manually by adversaries, 
as documented in~\cite{qin2020attacking}.
A complementary line of research is the design of domain-specific languages
for DeFi contracts, in the spirit of
the works~\cite{Arusoaie20arxiv,Biryukov17wtsc,EgelundMuller17bise,Thompson18isola} 
on languages for financial derivatives.
By leveraging primitives specifically tailored to DeFi,
these languages could simplify the task of analysing DeFi contracts:
actually, this task is overwhelmingly complex for current LP implementations,
which amount to thousands of lines of Solidity code.

% Existing LP platforms feature further DeFi instruments beyond loans,
% like \eg \emph{flash} loans \cite{aaveflashloan}. 
% We leave these advanced features for future investigation.

%% file: proofs.tex
\section{Supplementary material}

\begin{proofof}{lma: LP consistent minted token supply}
  The proof is by induction on the length of the trace
  from an initial state to the state $\tokBal \mid \LpS \mid \exchO$. 
  % on the execution(s) $\confG[0] \xrightarrow{}^* \confG = \tokBal \mid \LpS \mid \exchO$ that lead to configuration $\tokBal \mid \LpS \mid \exchO$.
  % 
  The base case is when $\tokBal \mid \LpS \mid \exchO$ is an initial configuration. Then, \eqref{eq: LP consistent minted token supply} trivially holds since $\minted[\LpS]$ is empty. 
  Now assume as induction hypothesis that~\eqref{eq: LP consistent minted token supply} holds for a reachable configuration $\confG$. We can show that the equation also holds for all configurations $\confG'$ such that $\confG \xrightarrow{\ell} \confG'$ by considering all possible cases for $\ell$.
  First note, that there are a number of cases where the state components involved in~\eqref{eq: LP consistent minted token supply} are not affected at all. These are: $\actTransfer{\pmvA}{\pmvB}{\valV:\tokT}$, $\exchUpdateOp$, $\accrueIntOp$, $\actBorrow{\pmvA}{\valV:\tokT}$, and $\actRepay{\pmvA}{\valV:\tokT}$.
  If  $\ell$ is $\actDeposit{\pmvA}{\valV:\tokT}$ or $\actRedeem{\pmvA}{\valV:\tokT}$ we note that the transition will increase and decrease both sides 
  of~\eqref{eq: LP consistent minted token supply} equally.
  Last, if  $\ell$ is $\actLiquidate{\pmvA}{\pmvB}{\valV:\tokT}{\valVi:\tokTi}$ or $\actTransferM{\pmvA}{\pmvB}{\valV:\tokT}$ the sum in the lhs of~\eqref{eq: LP consistent minted token supply} is kept constant (minted tokens are just transferred from one user to another one) and the rhs is not affected.
  \qed
\end{proofof}

\begin{proofof}{lma:ER-increasing}
  % The proof is by induction on the length of the trace
  % $\tokBal \mid \LpS \mid \exchO \xrightarrow{}^* \tokBali \mid \LpSi \mid \exchOi$. 
  % The base case is when $\tokBal \mid \LpS \mid  \exchO = \tokBali \mid \LpSi \mid \exchOi$. 
  % Then the lemma trivially holds since $\ER[\LpS]{\tokT} = \ER[\LpSi]{\tokT}$. 
  % % 
  % Now, assume as induction hypothesis that the lemma holds for all executions of length $n$. 
  % We are going to show that it also holds for executions of length $n+1$. 
  % In particular 
  We show that for a single transition $\tokBal \mid \LpS \mid \exchO \xrightarrow{\ell} \tokBali \mid \LpSi \mid \exchOi$ the following holds:
  \begin{enumerate}[(a)]
  \item $\ER[\LpS]{\tokT} < \ER[\LpSi]{\tokT}$,
    if $\ltsLabel = \accrueIntOp$ and $\exists\pmvA : (\lpBfun\,\pmvA)\,\tokT > 0$
  \item $\ER[\LpS]{\tokT} = \ER[\LpSi]{\tokT}$, otherwise. 
  \end{enumerate}
  Part (a) is easy to see, since the execution of $\accrueIntOp$ strictly increases the existing loans on $\tokT$ which are used in the numerator of the first case in~\eqref{eq: LP exchange rate}, without affecting the denominator. 
  
  For part (b) there are a number of cases where the state components involved in $ER$ are not affected at all. 
  These are: $\accrueIntOp$ (if $\not\exists\pmvA : (\lpBfun\,\pmvA)\,\tokT > 0$),   $\actTransfer{\pmvA}{\pmvB}{\valV:\tokT}$, $\exchUpdateOp$, and $\actTransferM{\pmvA}{\pmvB}{\valV:\tokT}$. 
  If $\ell$ is $\actBorrow{\pmvA}{\valV:\tokT}$, $\actRepay{\pmvA}{\valV:\tokT}$, or $\actLiquidate{\pmvA}{\pmvB}{\valV:\tokT[0]}{\tokT[1]}$ we note that the transition will increase and decrease the summands in the numerator of the first case in~\eqref{eq: LP exchange rate} equally.
  Last, If  $\ell$ is $\actDeposit{\pmvA}{\valV:\tokT}$ or $\actRedeem{\pmvA}{\valV:\tokT}$ we note that the transition will increase and decrease the numerator and denominator of the first case in~\eqref{eq: LP exchange rate} in quantities proportional to $\ER[\LpS]{}$. 
  \qed
\end{proofof}

\begin{proofof}{lma:constant-free-token-supply}
  The proof is by induction on the length of the trace
  $\tokBal \mid \LpS \mid \exchO \xrightarrow{}^* \tokBali \mid \LpSi \mid \exchOi$.
  The base case is when $\tokBal \mid \LpS \mid  \exchO = \tokBali \mid \LpSi \mid \exchOi$. Then the lemma trivially holds since $\supply{\tokBal,\LpS}(\tokT) = \supply{\tokBali,\LpSi}(\tokT)$. 
  Now, assume as induction hypothesis that the lemma holds for all executions of length $n$. We show that it also holds for executions of length $n+1$. In particular we show that for a single transition $\tokBal \mid \LpS \mid \exchO \xrightarrow{\ell} \tokBali \mid \LpSi \mid \exchOi$ token supplies remain constant by considering all possible cases for $\ell$,
  where state components of~\Cref{eq: free token supply} are affected. These are: $\actDeposit{\pmvA}{\valV:\tokT}$, $\actBorrow{\pmvA}{\valV:\tokT}$, $\actRedeem{\pmvA}{\valV:\tokT}$ and $\actLiquidate{\pmvA}{\pmvB}{\valV:\tokT}{\valVi:\tokTi}$. If $\ell$ is $\actDeposit{\pmvA}{\valV:\tokT}$, $\actBorrow{\pmvA}{\valV:\tokT}$ or $\actLiquidate{\pmvA}{\pmvB}{\valV:\tokT}{\valVi:\tokTi}$, changes applied to $\tokBal[\pmvA](\tokT)$ and $\lpFfun(\tokT)$ in~\Cref{eq: free token supply} cancel out. If $\ell$ is $\actRedeem{\pmvA}{\valV:\tokT}$, the same is true for user balance and lending pool balance of token $\uTok{\LpS}(\tokT)$.
  \qed
\end{proofof}

\begin{proofof}{lma:game0}
  By inspecting the formalization of the transitions it is easy to see which actions can increase or decrease in just one transition the net worth of a user on a specific token or in total, and to which extent.
  Indeed is is easy to see that the only actions that can modify the total net worth of a user $\pmvA$ are 
  $\accrueIntOp$, $\actTransfer{\pmvA}{\pmvB}{\valV:\tokT}$, $\actTransferM{\pmvA}{\pmvB}{\valV:\tokT}$, $\exchUpdateOp$ and $\actLiquidate{\pmvA}{\pmvB}{\valV:\tokT}{\valVi:\tokTi}$. Only the latter uses an action of the form ${\_}_{\pmvA}(\dots)$.  
  The rest of the actions of the form ${\_}_{\pmvA}(\dots)$ are 
  $\actDeposit{\pmvA}{\valV:\tokT}$, 
  $\actBorrow{\pmvA}{\valV:\tokT}$,  
  $\actRepay{\pmvA}{\valV:\tokT}$,  and  
  $\actRedeem{\pmvA}{\valV:\tokT}$. Inspecting their effect on wallets and loans we can notice that they simply exchange tokens in a way that keeps the net worth constant: $\actDeposit{\pmvA}{\valV:\tokT}$ and  $\actRedeem{\pmvA}{\valV:\tokT}$ simply swap amounts of free tokens and corresponding minted tokens proportionally to the exchange rate, while 
  $\actBorrow{\pmvA}{\valV:\tokT}$ and $\actRepay{\pmvA}{\valV:\tokT}$ simply move exact amounts of free tokens between wallets and loans. 
  \qed
\end{proofof}

\begin{proofof}{lma:game1}
  The main idea is that interest accrual affects net worth by increasing the share of deposited tokens (on which there is at least one non-empty loan) and increasing loan amounts. The only actions that increase a user's deposits are 
  $\actDeposit{\pmvA}{\valV:\tokT}$ and  
  $\actLiquidate{\pmvA}{\pmvB}{\valV:\tokT}{\valVi:\tokTi}$, 
  while the only action that decreases loans is 
  $\actRepay{\pmvA}{\valV:\tokT}$.
  \qed
\end{proofof}